\documentclass[prb,twocolumn,superscriptaddress,showpacs,floatfix]{revtex4}
\usepackage{graphicx}
\usepackage{color}
\usepackage{amsmath}
\usepackage{amssymb}
\usepackage{amstext}
\input{epsf}
\usepackage{latexsym}

\newcommand{\equ}[1]{Eq.~\ref{#1}}

\newcommand{\fig}[1]{Fig.~\ref{#1}}

\newcommand{\sect}[1]{Sec.~\ref{#1}}
\newcommand{\rem}[1]{}

\newcommand{\TT}{\mathcal{T}_3}

\newcommand{\virg}[1]{``{#1}''}

\newcommand{\abs}[1]{\left\vert#1\right\vert}

\newcommand{\To}{\longrightarrow}

\newcommand{\Ham}{\mathcal{H}}
\newcommand{\ket}[1]{|#1 \rangle}
\newcommand{\bra}[1]{\langle #1 |}

\newcommand{\matrice}[3]{\bra{#2} #1 \ket{#3}}

\newcommand{\pderiv}[2]{\frac{\partial \, #1}{\partial #2}}
\newcommand{\npderiv}[3]{\frac{\partial^#1 \, #2}{\partial #3^#1}}
\newcommand{\vett}[1]{\mathbf{#1}}
\newcommand{\trasf}[1]{\widetilde{#1}}

\newcommand{\varret}[3]{#1_{#3}(\vett{#2})}
\newcommand{\conf}[1]{\overrightarrow{#1}}
\newcommand{\traccia}[1]{\mbox{Tr}\left\{ #1 \right\}}
\newcommand{\Texp}[2]{\mathrm{T}_{#1} \exp \left( #2 \right)}

\begin{document}

\title{Phase Diagram of the Bose-Hubbard Model with $\TT$ symmetry}

\author{Matteo Rizzi}
\email{m.rizzi@sns.it} \affiliation{NEST CNR-INFM $\&$ Scuola Normale
Superiore, Piazza dei Cavalieri 7, 56126 Pisa, Italy}
\homepage{http://qti.sns.it}
\author{Vittorio Cataudella}
\email{cataudella@na.infn.it} \affiliation{COHERENTIA CNR-INFM $\&$
Dipartimento di Fisica, {\em Università Federico II}, 80126 Napoli,
Italy} \homepage{http://www.na.infn.it}
\author{Rosario Fazio}
\email{fazio@sns.it} \affiliation{NEST CNR-INFM $\&$ Scuola Normale
Superiore, Piazza dei Cavalieri 7, 56126 Pisa, Italy}
\homepage{http://qti.sns.it}
\date{\today}

\begin{abstract}
In this paper we study the quantum phase transition between the
Insulating and the globally coherent superfluid phases in the
Bose-Hubbard model with $\TT$ structure, the \virg{dice lattice}. Even
in the absence of any frustration the superfluid phase is characterized
by modulation of the order parameter on the different sublattices of
the $\TT$ structure. The zero-temperature critical point as a function
of a magnetic field shows the characteristic ``butterfly'' form. At
fully frustration the superfluid region is strongly suppressed.
In addition, due to the existence of the Aharonov-Bohm cages at
$f=1/2$, we find evidence for the existence of an intermediate
insulating phase characterized by a zero superfluid stiffness but
finite compressibility.	 In this intermediate phase bosons are
localized due to the external frustration and the topology of the $\TT$
lattice. We name this new phase the {\em Aharonov-Bohm} (AB) {\em
insulator}. In the presence of charge frustration the phase diagram
acquires the typical lobe-structure. The form and hierarchy of the Mott
insulating states with fractional fillings, is dictated	  by the
particular topology of the $\TT$ lattice.
The results presented in this paper were obtained by a variety of
analytical {\em methods: mean-field} and variational techniques to approach
the phase boundary from the superconducting side, and a strongly
coupled expansion appropriate for the Mott insulating region. In
addition we performed Quantum Monte Carlo simulations of the
corresponding (2+1)D XY model to corroborate the analytical
calculations with a more accurate quantitative analysis. We finally
discuss experimental realization of the $\TT$ lattice both with optical
lattices and with Josephson junction arrays.
\end{abstract}

\maketitle

\section{Introduction}
\label{intro} The Bose-Hubbard (BH) model~\cite{fisher89} is a paradigm
model to study a variety of strongly correlated systems as
superconducting films~\cite{sondhi97}, Josephson junction
arrays~\cite{fazio01} and optical lattices~\cite{jaksch98,greiner02}.
This model predicts the existence of a zero-temperature phase
transition from an insulating to a superfluid state which, by now, has
received ample experimental confirmation. The BH model is characterized
by two energy scales, an on-site repulsion energy between the bosons
$U$ and an hopping energy $t$ which allows bosons to delocalize. At
zero temperature and in the limit $U \gg t$ bosons are localized
because of the strong local interactions. There is a gap in the
spectrum for adding (subtracting) a particle{\em , hence} the compressibility
vanishes. This phase is named the Mott insulator. In the opposite limit,
$U \ll t$, bosons are delocalized and hence are in a superfluid phase.
There is a direct transition between
the Mott insulator and the superfluid state at a critical value of the
ratio $t/U$. This Superfluid-Insulator (SI) transition has been
extensively studied both theoretically and experimentally and we refer
to Refs. ~\onlinecite{sondhi97,fazio01,jaksch98,greiner02} (and
references therein) for an overview of its properties.

Magnetic frustration can be introduced in the BH-model by appropriately
changing the phase factors associated to the hopping amplitudes. The presence
of frustration leads to a number of interesting physical effects which has
been explored both experimentally and theoretically. In
Josephson arrays, where this is realized by applying an external
magnetic field, frustration effects have been studied extensively in the
past for both classical~\cite{frustration} and quantum systems~\cite{fazio01}.
Very recently a great interest in studying frustrated optical lattices has
emerged as well~\cite{jaksch03,sorensen05,polini05,osterloh05}. There are
already theoretical proposals to generate the required phases factors by
means of atoms with different internal states~\cite{jaksch03} or by applying
quadrupolar fields~\cite{sorensen05}.

The interest in the properties of dice lattices~\cite{sutherland86} has
been stimulated by the work by Vidal {\em et al.}~\cite{vidal98} on the
existence of localization, the so called Aharonov-Bohm (AB) cages, in
fully frustrated dice lattices without any kind of disorder. The
existence of these cages is due to destructive interference along all
paths that particles could walk on, when the phase shift around a
rhombic plaquette is $\pi$. Following the original paper by Vidal {\em
et al.}, several experimental~\cite{abilio99,naud01,serret03} and
theoretical
works~\cite{korshunov02,doucot02,cataudella03,korshunov03,protopopov04,bercioux04}
analyzed the properties of the AB cages. In the case of superconducting
networks most of the attention has been devoted to classical arrays
with the exception of Refs.~\onlinecite{doucot02,protopopov04} where a
frustrated quantum quasi-one dimensional {\em array} were studied.

In quantum arrays (charge) frustration can also be induced by changing
either the chemical potential (in optical lattices) or by means of a gate
voltage (in Josephson junction arrays).
This has the effect of changing the electrostatic energy needed to add/remove
a boson on a given island. The phase diagram present a typical lobe-like
structure~\cite{fisher89}. Moreover, depending on the range of the interaction,
it may also induce Wigner-like lattices of Cooper pairs commensurate with the
underlying lattice~\cite{bruder93}.
\begin{figure}
  \includegraphics[width=8cm]{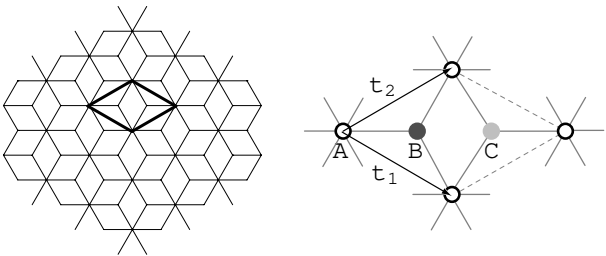}
\caption{
      The $\TT$ lattice: It consist of	hubs (with six nearest neighbours)
      connected to rims (three nearest neighbours). The $\TT$ structure is a
      Bravais	 lattice with a base inside the conventional unitary
      cell. The lattice vectors are $\vett{t_1}$ and $\vett{t_2}$. The
      basis is given by the sites A,B,C. Due to the fact that these sublattices
      are not self-connected and have different coordination numbers, we refer
   to this structure as tripartite. All rhombic plaquettes are identical,
   although differently oriented.}
\label{T3lat}
\end{figure}

The aim of this work is to study the phase diagram of a Bose-Hubbard
model on a $\TT$ lattice (shown in Fig.\ref{T3lat}). We will consider
both the cases of electric and magnetic frustration. The location and
the properties of the phase diagram will be analyzed by a variety of
approximate analytical methods (mean-field, variational Gutzwiller
approach, strong coupling expansion) and by Monte Carlo simulations.
The $\TT$ lattice has been experimentally realized in Josephson
{\em arrays~\cite{serret03}}. In addition we show that it is possible to
realize it experimentally also with optical lattices. Although the main
properties of the phase diagram are common to both experimental
realizations, there are some differences which are worth to be highlighted.

The plan of the paper is the following. In the next Section we will
discuss the appropriate model for both the case of a Josephson junction
array and optical lattices~(\sect{models}) and discuss in some detail
how the $\TT$ structure can be realized in an optical
lattice~(\sect{optical}). In the same Section we introduce the relevant
notation to be used in the rest of the paper. A description of the
various analytical approaches used to obtain the phase diagram will be
given in~\sect{analytic}. The zero-temperature phase diagram, in the
presence of magnetic and electric frustration, will then be described
in~\sect{diagram}. We first discuss the unfrustrated case and
afterwards we consider the role of electric and magnetic frustration
respectively. Due to the particular topology of the $\TT$ lattice the
superconducting phase is  characterized, even at zero frustration, by a
modulation of the order parameter on the different sublattices (i.e.
hubs and rims), which indicates a different phase localization on
islands depending on their coordination number. A uniform electrostatic
field gives rise to a lobe structure in the phase diagram which is
discussed for the $\TT$ array {\em in~\sect{electric}. The effect of a
uniform external magnetic field, discussed in~\sect{magnetic}, may induce
important qualitative changes in the phase diagram in the
case of fully frustration.} In particular we will provide evidences that there
is an important signature of the Aharonov-Bohm cages in the quantum
phase diagram.	 It seems that	  due to the AB cages a new phase
intermediate between the Mott insulating and superfluid phases should
appear. On varying the ratio between the hopping and the Coulomb energy
the system undergoes {\em two} consecutive quantum phase transitions.
At the first critical point there is a transition from a Mott insulator
to a {\em Aharonov-Bohm insulator}. The stiffness vanishes in both
phases but the compressibility is finite only in the Aharonov-Bohm
insulator. At a second critical point the system goes into a superfluid
phase. Most of the analysis is presented by using approximated
analytical methods. These results will be checked against Monte Carlo
simulations that we present in~\sect{QMC methods}. Few details of the
mapping of the model used in the simulation are reviewed in the
Appendix~\ref{3dxymap}. The concluding remarks are summarized
in~\sect{discuss}.

\section{Quantum Phase Model on a $\TT$ array}
\label{models}

Both Josephson arrays and optical lattices are experimental
realizations of the Bose-Hubbard model
\begin{eqnarray}
        \mathcal{H} & = & \mathcal{H}_U+ \mathcal{H}_t \nonumber \\
      & = & \frac{1}{2}\sum_{ij} (n_i-n_0) \mathcal{U}_{i,j} (n_i-n_0)
      \nonumber  \\
      & - & \frac{\tilde{t}}{2} \sum_{<ij>} (b^{\dag}_i	b_j + h.c.) \;\; .
\label{hubbard}
\end{eqnarray}

When the mean occupation ${\bar n}$ on each lattice site is large, one
is allowed to introduce the phase operator $\varphi_i$ by approximating
the boson annihilation operator on site $i$ by $b_i \simeq \sqrt{{\bar
n}}~\exp{[\imath \varphi_i]}$. The density $n_i$ and phase $\varphi_i$
operators are canonically conjugate on each site
\begin{equation}
         \left[ n_{i}, e^{\pm \imath
         \varphi_{j}} \right] = \pm \delta_{i,j} \,
         e^{\pm \imath \varphi_{i}} \, .
\label{Heisen}
\end{equation}

In the present work we will focus our attention on the quantum rotor
version of the model in Eq.(\ref{hubbard}) that reads:
\begin{eqnarray}
        \mathcal{H} & = & \frac{1}{2}\sum_{ij} (n_i-n_0)
        \mathcal{U}_{i,j} (n_j-n_0) \nonumber  \\
        & -& t \sum_{<ij>}\cos (\varphi_i- \varphi_j- A_{i,j}) \;\; .
\label{hamiltonian}
\end{eqnarray}
The first term on the r.h.s of Eq.(\ref{hamiltonian}) represents the
repulsion between bosons ($\mathcal{U}_{i,j}$ depends on the range of
the interaction and on its detailed form). The second term is due to
the boson hopping ($t=\bar{n}\tilde{t}$ is the coupling strength)
between neighboring sites (indicated with $\langle . \rangle$ in the
summation). The gauge-invariant definition of the phase in presence of
an external vector potential $\vett{A}$ and flux-per-plaquette $\Phi$
($\Phi_0 = h \, c / \, 2 \, e$ is the flux quantum) contains the term
$ A_{i,j} = \frac{2 \pi}{\Phi_0} \int_{i}^{j} \vett{A} \cdot d\vett{l}
\; . $ All the observables are function of the frustration parameter defined as
\begin{equation}
         f = \frac{1}{\Phi_0}
         \int_{P} \vett{A} \cdot d\vett{l} = \frac{1}{2 \pi} \sum_P \, A_{i,j}
\end{equation}
where the line integral is performed over the elementary plaquette. Due
to periodicity of the model it is sufficient to consider values of the
frustration $0 \le f \le 1/2$. Charge frustration is due to a
non-integer value $n_0$. As for the magnetic frustration also in this
case the properties will be periodic under the transformation $n_0
\rightarrow n_0 +1$. Due to the additional symmetry $n_0 \rightarrow
-n_0$ it is sufficient to consider value of the charge frustration
$n_0$ in $[0,1/2]$. Differently from the magnetic frustration the
value of $n=1/2$ does not necessarily correspond to fully (charge)
frustration as this depends on the range of the interaction
$\mathcal{U}_{i,j}$.

The $\TT$ lattice~\cite{sutherland86} is represented in Fig.\ref{T3lat},
the lines between the sites corresponds to those links where boson hopping
is allowed.
The $\TT$ structure is not itself a Bravais lattice, but could be considered as
a lattice with a base inside the conventional unitary cell (see \fig{T3lat})
defined by the vectors
\begin{eqnarray}
         \vett{t_1} &=& \left(3/2; -\sqrt{3}/2\right) a \nonumber \\
         \vett{t_2} &=& \left(3/2; +\sqrt{3}/2\right) a \;\; . \nonumber
\end{eqnarray}
where $a$ is the lattice constant. The lattice sites of
the base are at positions
$$
   \vett{d_A} = \left(0; 0\right) a
   \hspace{0.3cm}
   \vett{d_B} = \left(0; 1\right) a \hspace{0.3cm}
   \vett{d_C} = \left(0; 2\right) a \;\; .
$$
The reciprocal lattice ($\vett{g_a} \cdot \vett{t_b} = 2 \pi \delta_{a,b}$)
vectors are defined as
\begin{eqnarray}
       \vett{g_1} &=& \frac{2\pi}{a}\left(1/3;-\sqrt{3}/3\right) \nonumber \\
     \vett{g_2} &=& \frac{2\pi}{a}\left(1/3;+\sqrt{3}/3\right) \;\; . \nonumber
\end{eqnarray}
In several situations  it turns out to be more convenient to label the
generic site $i$ by using the position of the cell $\vett{t} = n_1
\vett{t_1} + n_2 \vett{t_2}$ ($-N_l \leq n_l < N_l$) and the position
within the cell $\alpha = A,B,C$. In the rest of the paper we
either use the index $i$ or the pair of labels $(\vett{t},\alpha)$. A
generic observable $W_i$ can be written henceforth as
$\varret{W}{t}{\alpha}$. By imposing Born-Von Karman periodic boundary
conditions its Fourier transform is given by
\begin{equation}
       \varret{\trasf{W}}{K}{\alpha} =
       \frac{1}{\sqrt{4 N_1 N_2}} \sum_{\vett{t}} \varret{W}{t}{\alpha}\
       e^{-\imath \vett{K}\cdot\vett{t}}
\label{fourier}
\end{equation}
with $\vett{K} = k_1 \vett{g_1} + k_2 \vett{g_2}$ in the first
Brillouin zone.

It is also useful to introduce a connection matrix $\mathcal{T}$ whose
entries are non-zero only for islands connected by the hopping.
More precisely	 $\varret{T}{t,t'}{\alpha,\gamma}=1$ if site
$\alpha$ of cell $\vett{t}$ is connected by a line (see
Fig.\ref{T3lat}) to site $\gamma$ of cell $\vett{t'}$ and
$\varret{T}{t,t'}{\alpha,\gamma}=0$ otherwise. The local coordination
number is thus defined as $z_{\alpha} = \sum_{\vett{t'},\gamma}
\varret{T}{t,t'}{\alpha,\gamma}$. It is $z=6$ for the hubs (labelled by A)
and $z=3$ for the rims (labelled by B and C). For later convenience we also
define the matrix $\mathcal{P}$ with elements
\begin{equation}\label{Pdef}
       \varret{P}{t,t'}{\alpha,\gamma}\ =
       \varret{T}{t'-t}{\alpha,\gamma}\
       e^{\imath \varret{\vett{A}}{t,t'}{\alpha,\gamma}}
\end{equation}
which includes the link phase factors which appear if the system is
frustrated. In the whole paper we fix $k_B=\hbar=c=1$.

In the next two subsections we give a brief description of the origin
and characteristics of the coupling terms in the Hamiltonian of
Eq.(\ref{hamiltonian}) both for Josephson and optical arrays. In
addition we show how to realize optical lattices with $\TT$ symmetry.

\subsection{Josephson junction arrays}
\label{jjas}

Since the first realization of a Josephson Junctions Array
(JJA)~\cite{IBM}, these systems have been intensively studied as ideal
model systems to explore a wealth of classical
phenomena~\cite{classical} such as phase transitions, frustration
effects, classical vortex dynamics and chaos. One of the most
spectacular result was probably the experimental
observation~\cite{BKTexp} of the Berezinskii-Kosterlitz-Thouless
transition (BKT)~\cite{BKTth}. Indeed, well below the BCS transition
temperature, and in the classical limit, JJAs are experimental
realization of the XY model. For sufficiently small (submicron) and highly
resistive (normal state resistance $R_N  > R_Q = h /4 e^2$) junctions
quantum effect start to play an important role. In addition to the
Josephson energy, which controls the Cooper pair tunnelling between
neighboring grains, also the charging energy $e^2 / 2 C$ ($C$ is the
geometrical junction capacitance) becomes important.

Experiments on JJAs are performed well below the BCS critical
temperature and thus each island is in the superconducting state. The
only important dynamical variable is the phase $\varphi_{i}$ of the
superconducting order parameter in each island, canonically conjugated
to the number of extra Cooper pairs $n_{i}$ present on that island. In
Eq.(\ref{hamiltonian}), the coupling constant $t$ equals the Josephson
coupling. Hence the second term in Eq.(\ref{hamiltonian}) represent
the Josephson energy. The first term is due to charging energy which
can be evaluated by assuming that each island  has a capacitance to
the ground $C_0$ and each junction a geometrical capacitance $C$. The
electrostatic interaction between the Cooper pairs is defined as
\begin{equation}
\label{Uij}
         \mathcal{U} = 2 e^2 \mathcal{C}^{-1} \; .
\end{equation}
The capacitance matrix is given by
\begin{equation}
\label{Cij}
         C_{i,j} = (C_0 + z_{i} C) \, \delta_{i,j} - C \, T_{i,j} \; .
\end{equation}
Since both the connection and the capacitance matrices depend only on the
distance between the cells (and on the base index of both sites), their space
dependence can be simplified to
\begin{equation}
         \varret{C}{t,t'}{\alpha,\gamma} = \varret{C}{0,t'-t}{\alpha,\gamma}
      \equiv \varret{C}{t'-t}{\alpha,\gamma}
\end{equation}

An estimate of the range of the electrostatic interaction is given
by~\cite{mooij} $\lambda \approx \sqrt{C / C_0}$. The charge frustration
$n_0$, that we assume to be uniform, can be induced by an external
(uniform) gate voltage $V_0 = n_0 / C_0$.

\begin{figure}
  \includegraphics[width=9cm]{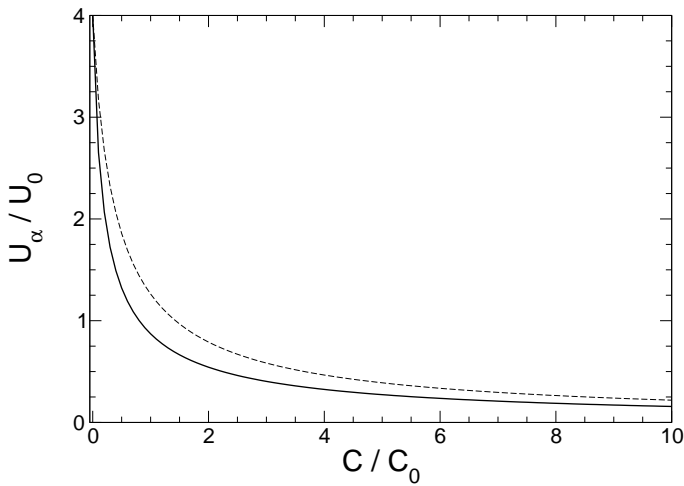}
\caption{
   Electrostatic energy (in units of $U_0 = e^2 / 2 C_0$) required
   to put an extra Cooper pair (for zero external charge) on an hub
   (straight line) and on a rim (dashed) as a function of the reduced
   capacitance $C/C_0$}
\label{ediff}
\end{figure}

Due to the particular structure of the $\TT$ lattice, the charging
energy of a single (extra) Cooper pair placed on a given islands {\it
depends} on that site being a rim or a hub as shown in
Fig.\ref{ediff}. As a consequence quantum fluctuations of the phase of
the superconducting order parameter may be different in the two
different cases (rims or hubs). We will see in Sec.\ref{diagram0} that
this property is responsible for an additional modulation of the order
parameter in the superconducting phase.

\subsection{Optical lattices}
\label{optical}

Following the work of Jaksch {\em et al.}~\cite{jaksch98}, optical
lattices have been widely studied as concrete realization of the
Bose-Hubbard model that is, as we saw, directly related to the quantum
phase model studied in this paper. The experimental test of the SI 
transition~\cite{greiner02} has finally opened
the way to study strongly correlated phenomena in trapped cold atomic
gases. Very recently, several works addressed the possibility to
induce frustration in optical
lattices~\cite{jaksch03,sorensen05,polini05,osterloh05}. It is
therefore appealing to test the properties of the $\TT$ lattice also
with optical lattices once it is known how to create $\TT$ lattices by
optical means.

\begin{figure}[htbp]
  \includegraphics[width=8cm]{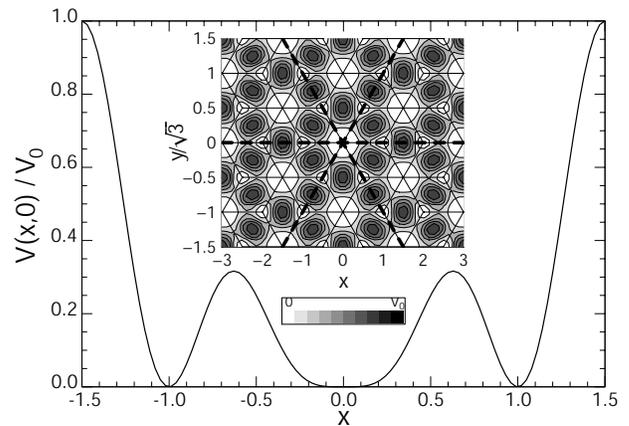}
\caption{
   Optical potential with $\TT$ symmetry generated by three counter-propagating
   laser beams. The inset shows the bidimensional contour plot while in the
   figure the details of the profile along a line connecting three sites
   (placed at positions $x=1$, $x=0$ and $x=-1$) is shown. The sites $x=-1,1$
   are rims while the site at $x=0$ is a hub. Also here, as in the case of
   JJAs, the different form of the potential implies that the on-site energy
   $U_0$ is different for hubs and rims.}
\label{figopt}
\end{figure}

Here we propose an optical realization of a $\TT$ structure by means of
three counter-propagating pairs of laser beams. These beams divide the
plane in six sectors of width $60$° (see the inset of Fig.\ref{figopt})
and are linearly polarized such to have the electrical field in the
$xy$ plane. They are identical in form, apart from rotations, and have
wavelength equal to $\lambda = 3/2 \, a$ ($a$ is the lattice constant.
Given a polarization of a pair of lasers on the $y$-axis $\vec{E_1} =
(0,E_{y})$ the other two pairs are obtained by rotating $\vec{E_1}$ of
$120$° and	$-120$° around the $z$-axis. The square modulus of the
total field gives rise to the desired optical potential as it is shown
in Fig.\ref{figopt}.

The form of the potential landscape also in this case imposes that the
on-site repulsion may be different for hubs and rims.	 It is however
diagonal
\begin{equation}
   \mathcal{U} = U_r \; \mathcal{I}_{r} + U_h \;  \mathcal{I}_{h} \;\; .
   \label{onsite}
\end{equation}
The subscript $h,r$ denotes the respectively the hub and rim sites and 
$\mathcal{I}_{h,r}$ are the projectors on the corresponding 
sublattices. In Eq.(\ref{hamiltonian}), now, the coupling $t$ 
describes the hopping amplitudes for bosons, $n_0$ is proportional to 
the chemical potential, $A_{ij}$ is the effective ``magnetic 
frustration'' that in this case may have several different origin 
depending on the scheme used. For simplicity we will always refer to 
$\vec{A}$ as to the vector potential and we will use the magnetic 
picture also for optical lattices.

\section{Analytic approaches}
\label{analytic}

The SI transition has been studied by a variety of methods; here we
apply several of them to understand the peculiarities that emerge in
the phase diagram due to the $\TT$ lattice structure. The results that
derive from these approaches will be presented in the next section.

The location of the critical point depends on the exact form and the
range of $\mathcal{U}_{i,j}$. This issue is particulary interesting
when discussing the role of electric frustration. In the paper we
address the dependence of the phase boundary on the range of the
interaction in the mean-field approximation. The variational Gutzwiller
ansatz and the strong coupling expansion will be analyzed only for the
on-site case of Eq.(\ref{onsite}). In the case of magnetic frustration
the form of $\mathcal{U}_{i,j}$ leads only to quantitative changes so,
also in this case, we discuss only the on-site case.

\subsection{Mean field approach}
\label{scmf}

The simplest possible approach to study the SI phase boundary consists in
the evaluation		of the superconducting order parameter, defined as
$$
      \psi_i = \langle e^{-\imath
      \varphi_i} \rangle \, ,
$$
by means of a mean-field approximation. By neglecting terms quadratic
in the fluctuations around the mean field value, the	hopping part of
the Hamiltonian can be approximated as
\begin{eqnarray}
\label{HJmf}
        \nonumber
         \Ham_{t}^{(mf)} = - \frac{1}{2} &t& \sum_{i,j}
        e^{-\imath \varphi_i(\tau)}\ P_{i,j}\
        \psi_j + h.c.
\end{eqnarray}
The order parameter is then determined via the self-consistency
condition
\begin{equation}
\label{eq:scmf}
        \psi_i (\tau') =
          \frac{\traccia{e^{\imath \varphi_i(\tau')} \,
        e^{- \beta \, \Ham_{U}} \,
        \Texp{\tau}{\int_0^{\beta} \Ham_{t}^{(mf)}(\tau)}}}
        {\traccia{ e^{- \beta \, \Ham_{U}}
        \Texp{\tau}{\int_0^{\beta} \Ham_{t}^{(mf)}(\tau)}}} \; .
\end{equation}
In the previous equation, $\mathrm{T}_{\tau}$ is the time-ordering in
imaginary time $\tau$ and $\beta = 1/T$ . The $\tau$ dependence of the
operators is given in the interaction representation $W(\tau) = e^{\tau \,
\Ham_{U}} \, W \, e^{-\tau \, \Ham_{U}}$. For simplicity we already assumed
the order parameter independent on the imaginary time. One can indeed verify
that this is the case in the mean-field approximation. Close to the phase
boundary the r.h.s. of \equ{eq:scmf} can be expanded in powers of the order
parameter and the phase boundary is readily determined.

A central quantity in the determining the transition is
the phase-phase correlator
\begin{equation}
\label{corrG}
      G_{i,j}(\tau)	 =
      \langle \mathrm{T}_{\tau} e^{\imath \phi_i(\tau)}
      e^{- \imath \phi_j(0)}\rangle_{U}
\end{equation}
where the average is performed with the charging part of the
Hamiltonian only. Charge conservation imposes that the indexes $i, j$
are equal. The Matsubara transform at $T=0$ of the correlator reads
\begin{equation}
\label{Gtrasf}
      \trasf{G}_{i,i}(\omega) = \int_{-\infty}^{\infty} G_{i,i} (\tau)
        e^{\imath \, \omega \, \tau }
      = \sum_{s=\pm}
         \frac{1}{\Delta E_{\alpha,s} - \imath s \omega}
\end{equation}
where $\Delta E_{\alpha,\pm}$ are the excitation energies (for zero
Josephson tunnelling) to create a particle (+) or a hole (-) on a site
of the sublattice $\alpha$ where $i$ lies.

In the case of the $\TT$ lattice considered here even at
zero magnetic field the order parameter is not uniform. The tripartite nature
of the lattice results in a vectorial mean field $\psi$ with one component
for each sublattice. In the general case the linearized form of
Eq.(\ref{eq:scmf}) can be rewritten as
\begin{equation}
      \varret{\psi}{t}{\alpha} = \frac{t}{2}
         \sum_{\gamma} \sum_{\vett{t'}}
         \trasf{G}_{\alpha,\alpha}(0)
         \varret{P}{t,t'}{\alpha,\gamma}
         \varret{\psi}{t'}{\gamma}
\end{equation}
that, due to the topology of the lattice is equivalent to
\begin{equation*}
   \varret{\psi}{t}{\alpha} = \frac{t^2}{4}
         \trasf{G}_{A,A}(0) \,
         \trasf{G}_{B,B}(0) \,
         \sum_{\gamma} \sum_{\vett{t'}}
         \varret{P^2}{t,t'}{\alpha,\gamma}
         \varret{\psi}{t'}{\gamma}
\end{equation*}

\begin{figure}[htbp]
  \includegraphics[width=4cm]{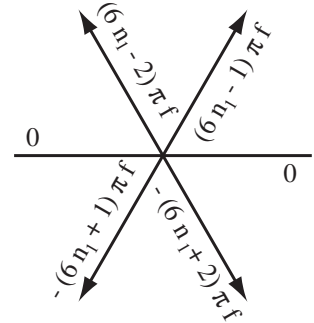}
  \caption{Magnetic phase pattern with the gauge choice
    $\vett{A}~=~(x~-~\sqrt{3}~y)~\frac{2 \Phi_0}{\sqrt{3}
      a^2}~f~\vett{\hat{y}}$}.
  \label{fig:pattern}
\end{figure}

The phase transition is identified with a non-trivial solution to this
secular problem, i.e. one should determine $\pi_{max}$, the largest
eigenvalue of $P$. This requirement
translates in the following equation for the critical point
\begin{equation}\label{eq:ejcrit}
       t_{cr} = 2
       \frac{\pi_{max}^{-1}}
       {\sqrt{\trasf{G}_{A,A}(0) \trasf{G}_{B,B}(0)}}
\end{equation}
In deriving the previous equation we used the fact that the sites $B$
and $C$ in the elementary cell (see Fig.\ref{T3lat}) have the same
coordination number and therefore the phase-phase correlator is the
same. In addition to the evaluation of the Matsubara transform at zero
frequency of the phase correlator, one has to determine the eigenvalues
of the gauge-link matrix $\mathcal{P}$. With a proper gauge choice it
is possible to reduce this matrix to a block diagonal form. For
rational values of the frustration, $f = p / q$, by choosing $\vett{A}
~=~(x~-~\sqrt{3}~y)~\frac{2 \Phi_0}{\sqrt{3} a^2}~f~\vett{\hat{y}}$,
the magnetic phase factors $\varret{\vett{A}}{t,t'}{i,j}$ (shown in
\fig{fig:pattern}) have a periodicity of $r \times 1$ elementary cells
with $r = LCM(q,3)/ 3$. This implies that in the Fourier space (see
Eq.(\ref{fourier})) the component $k_2$ is conserved and that $k_1$ is
coupled only with the wavevectors $k_1^{(m)} = k_1 + 2 \pi m/ r$
($m=0,...r-1$).
The determination of $\pi_{max}$ is therefore reduced to the
diagonalization of a $3r \times 3r$ matrix
($\trasf{P}_{\alpha,\gamma}(k_1)$ is $r \times r$)
\begin{equation}
       \trasf{\mathcal{P}}(k_1,k_2) = \delta_{k_2,0}
   \left(
       \begin{array}{ccc}
       0 & \trasf{P}_{A,B}(k_1) & \trasf{P}_{A,C}(k_1) \\
       \trasf{P}^{\dag}_{A,B}(k_1) & 0 & 0 \\
       \trasf{P}^{\dag}_{A,C}(k_1) & 0 & 0 \\
       \end{array} \right)
\end{equation}
with $(k_1,k_2)$ belonging to the reduced Brillouin zone $\mathbb{B}_r
= \{ 0 \leq k_i < 2 \pi / r \}$.

The matrix $\mathcal{P}$ has $r$ zero eigenvalues, and $r$ pairs of
eigenvalues equal in absolute value given by the reduced secular
equation
\begin{equation*}
 \left[ \trasf{P}_{1,2}(k_1) \trasf{P}^{\dag}_{1,2}(k_1) + \trasf{P}_{1,3}(k_1)
       \trasf{P}^{\dag}_{1,3}(k_1) \right] \ \trasf{v}_1 = \pi^2 \ \trasf{v}_1
\end{equation*}
This simplification allows us to deal with $r \times r$ matrices instead
of $q \times q$.

The inclusion of a finite range interaction, important only for Josephson
arrays, leads to a richer lobe structure in presence of electrostatic
frustration. The calculation of the lobes will be done within the mean field
theory only.

\subsection{Gutzwiller variational approach}
\label{Gutz}

A different approach, still mean-field in spirit, that allows to study
the properties of the superconducting phase is the Gutzwiller variational
ansatz adapted to the Bose-Hubbard model by Rokhsar and Kotliar~\cite{rokhsar}
The idea is to construct a variational wave-function
for the ground state starting from the knowledge of the wave-function in
the absence of the interaction term $\mathcal{H}_U$ in the Hamiltonian.
In this case, and in absence of magnetic frustration,
the ground state has all the phases aligned along a fixed direction
$\theta$. In the boson number representation it reads
\begin{equation}
         \ket{GS}_{U=0} =
        \sum_{\{n_i\}} e^{\imath \sum_{i}n_i \theta} \ket{\{n_{i}\}}
\end{equation}
A finite charging energy, tends to suppress the components of the state
with large charge states, a variational state can then be constructed
through the ansatz
\begin{equation}
        \ket{GS} = \sum_{\{n_i\}} c_{n_1,\cdots,n_N} \ket{\{n_i\}}
\end{equation}
where
\begin{equation}
         c_{\{n_i\}} = \frac{1}{\sqrt{N_{GS}}} e^{\imath \sum_{i} n_i \theta}
         e^{- \sum_{i} \ \frac{K_{i}}{2}
         (n_{i} - \overline{n}_{i})^2} \, .
\label{gutz}
\end{equation}
In Eq.(\ref{gutz}) $N_{GS}$ is a normalization factor and $K_{i}$ and
$\overline{n}_{i}$ are variational parameter to be determined by
minimizing the ground state energy. The Mott insulator is characterized
by $K		= \infty$, i.e. by perfect localization of the charges, $K =
0$ is the limit of zero charging, a finite value of $K$ describes a
superfluid phase where the phase coherence has been established albeit
suppressed by quantum fluctuations.

\subsection{Strong coupling perturbation theory}
\label{pert}

Both methods illustrated in Sections \ref{scmf} and \ref{Gutz} are
based on the analysis of the superconducting phase and on the
determination of the phase boundary as the location of points where the
superfluid order parameter vanishes. A complementary approach, which
analyzes the phase boundary from the insulating side, was developed by
Freericks and Monien~\cite{freericks}. The method was applied to the
case of square and triangular lattices in Ref.~\onlinecite{niemeyer} for the
Bose-Hubbard model and in Ref.~\onlinecite{kim} for the quantum rotor model.
In this section we describe how to adapt the method to the $\TT$
lattice. We will present the results of this analysis, particularly
important for the fully frustrated case, in~\sect{magnetic}.

In the insulating phase the first excited state is separated by the
ground state by a (Mott) gap. In the limit of vanishing hopping
the gap is determined by the charging energy needed to place/remove
an extra boson at a given lattice site. The
presence of a finite hopping renormalizes the Mott gap which,
at a given critical value, vanishes. The system becomes compressible,
and the bosons, since are delocalized, will condense onto a superfluid
phase. It is worth to emphasize that the identification of the SI
boundary with the point at which the gap vanishes is possible as the
bosons delocalize once the energy gap is zero. As we will see, in the
case of $\TT$ lattice the situation becomes more complex. In the
presence of external magnetic frustration it may happen that though
the Mott gap is zero, the states are localized and therefore the
charges cannot Bose condense. In this cases between the Mott and
superconducting region an additional compressible region (with zero
superfluid stiffness) may appear. In order to keep the expressions as
simple as possible we consider only the case of on-site interaction,
though we allow a different $U$ for hubs and rims as in
Eq.(\ref{onsite}). The possible existence of such a phase, however,
does not depend on the exact form of $\mathcal{U}_{i,j}$.  The strong
coupling expansion is particularly useful for $\TT$ lattice as it may
help in detecting, if it does exist, the intermediate phase.

In the strong-coupling approach of Freericks and Monien the task is to
evaluate, by a perturbation expansion in $t/U$, the energy of the
ground and the first excited state in order to determine the point
where the gap vanishes. We denote the ground and first excited levels
by $E_M^{gs}$ and $E_M^{exc}$ respectively. The choice of the starting
point for the perturbation expansion is guided by the nature of the
low-lying states of the charging Hamiltonian. When $n_{0} < 1/2$ (and
in zero-th order in $t/U$) the ground state of the electrostatic
Hamiltonian is ($n_{i} = 0 \; \forall i$) and first excited level is
given by a single extra charge localized on a site. Levels
corresponding to charging a hub and a rim are nearly degenerate (i.e.
$(U_r-U_h)/(U_r + U_h) \ll 1$, with the hub being lower in energy). As
the strength of the hopping is increased, the insulating
gap decreases.
We would like to stress, and this is an important difference emerging
from the $\TT$ topology, i.e. the location of the extra charge (on a
hub or a rim) requires a different energy. This in turn has important
consequences in the structure of the perturbation expansion. 

Up to the second-order in the tunnelling, the ground state energy at $n_0 = 0$
is given by
\begin{equation}
\label{ground}
      E_{M}^{gs} = - \frac{2 \cdot 2 N}{(U_h + U_r)/2} \; \frac{t^2}{4}
\end{equation}
where $N$ is the number of sites and $2 N$ the number of hub-rim links
in the lattice. Note that the first-order correction vanishes because
the tunnelling term does not conserve local number of particles.

Due to nearly degeneracy of the excited levels, one is not allowed to
perturb each of them independently but has to diagonalize the zeroth
and the first order terms simultaneously. One has to
diagonalize the following matrix:
\begin{equation}
      \mathcal{Q}^{(1)} = \frac{1}{2} \mathcal{U} -
      \frac{t}{2} \mathcal{P}
\label{first}
\end{equation}
This task can be reduced to the diagonalization of a $3 r(f) \times 3
r(f)$ matrix with a proper choice of the gauge (see Section~\ref{scmf}).

For example, the (degenerate) lowest eigenvalue at $f=1/2$ is
\begin{equation}
      \left.Q_{min}^{(1)}\right|_{f=1/2} = \frac{U_h+U_r}{4}
      - \frac{1}{2}\, \sqrt{6 t^2 + \left(\frac{U_r - U_h}{2}\right)^2}
\end{equation}
which reduces to $U/2 - t \, \sqrt{6}/2$ in the case of perfectly
degenerate charging energy. It must be stressed that all the energy
bands are flat, independently of the values of the charging energies
(it depends only on the peculiar $\mathcal{P}$ structure).

\begin{figure}
  \centerline{  \includegraphics[width=7cm]{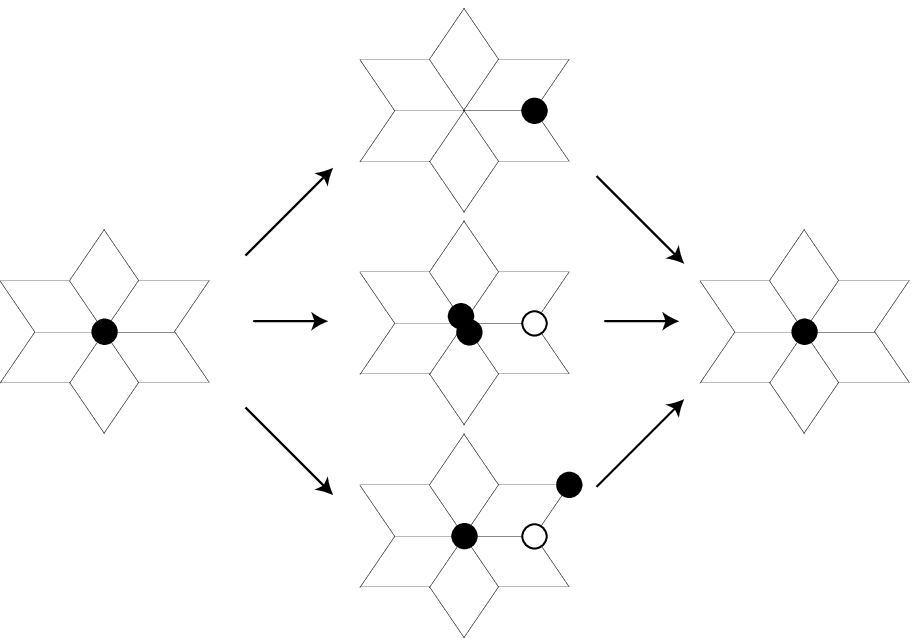}}
   \hspace{1cm}
\centerline{  \includegraphics[width=7cm]{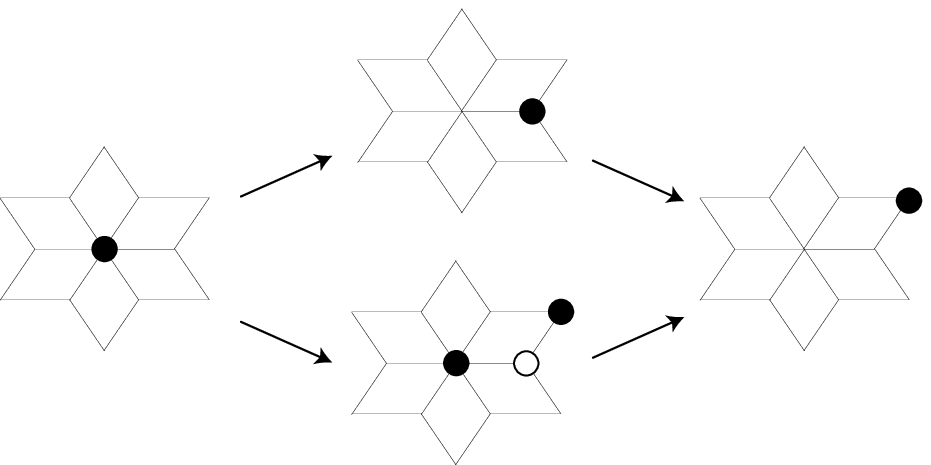}}
\caption{
   Intermediate charge states involved in the definition of
      Eq.(\ref{second}). In the upper panel the contributions to the diagonal
   part are shown while in the lower panel there are the
      contributions to the off-diagonal part. The processes represented here
   are those contributing to the second order in the hopping
   amplitude. The black/white circles represent one extra +/- Cooper
   pair on a given site.}
\label{intermediate}
\end{figure}

The second order perturbation term should be calculated on the lowest
energy manifold: moreover only matrix elements between states of the
same manifold are allowed. Nonetheless, it is simpler to write the
different contributions in the usual basis of hub and rims (see
Fig.(\ref{intermediate})). The first excited state, to second order in
tunnelling is given by
\begin{equation}
\label{excited}
      E_{M}^{(1)} = Q_{min}^{(1)} + \frac{t^2}{4} Q_{min}^{(2)}
\end{equation}
where $\mathcal{Q}^{(2)}$ is the second order matrix and can be split
into separate sub-matrices on different sub-lattices, i.e.
\begin{equation}
      \mathcal{Q}^{(2)} = \mathcal{Q}^{(2)}_h \, \mathcal{I}_h +
      \mathcal{Q}^{(2)}_r	 \, \mathcal{I}_r
\end{equation}
Such a decomposition is possible because after two tunnelling events
the boson come back to the initial sublattice.
\begin{eqnarray}
      \mathcal{Q}(2)_{h} & = &
         z_h \frac{\mathcal{I}_h}{(U_h-U_r)/2} +
      z_h \frac{\mathcal{I}_h}{\left(U_h-(4 U_h + U_r)\right)/2}
      \nonumber \\
         &+&  \left( 2 \cdot 2N - 2 z_h \right) \frac{\mathcal{I}_h}
      {(U_h-(2 U_h + U_r))/2}
         \nonumber \\
         & + &
         \frac{\mathcal{P}^2 - z_h \mathcal{I}_h}{(U_h-U_r)/2}
      + \frac{\mathcal{P}^2 - z_h \mathcal{I}_h}
      {\left(U_h-(2 U_h + U_r)\right)/2}
\label{second}
\end{eqnarray}
($\mathcal{Q}(2)_{r}$ is defined in a similar way) where
$\mathcal{I}_{h,r}$ are the projectors on the hub and rim sublattices.
After some algebra
and by changing basis to the one composed by the eigenvectors of
Eq.(\ref{first}), one gets the first excited energy level. The task is
now to determine the location of points at which the gap, given by the
difference of Eq.(\ref{excited}) and Eq.(\ref{ground}), vanishes. It
is worth to stress that the thermodynamically divergent contributions
wash out exactly their analogous in the ground state expression of
Eq.\ref{ground}.

We discuss the results deriving from this approach in the next Section where we
analyze the phase diagram.

\section{Phase diagram}
\label{diagram}

In order to keep the presentation as clear as possible we first
discuss the main features of the phase diagram by means of the
analytical approaches introduced before. We will then corroborate these
results in a separate section by means of the Monte Carlo simulations.

The value of the critical Josephson coupling as a function of the range
of the electrostatic interaction, in the absence of both electric and
magnetic frustration is discussed first. The effect of frustration,
either electric or magnetic will then be discussed in two separate
sections. In the case of electrical frustration the topology of a $\TT$
lattice gives rise to a rather rich lobe structure, the overall picture
is nevertheless very similar to the one encountered in the square
lattice. Much more interesting, as one would suspect, is the behaviour
of the system as a function of the magnetic frustration. The location
of the phase boundary shows the characteristic butterfly shape
with an {\em upturn} at fully frustration typical of the
$\TT$. In addition, at $f=1/2$, a very interesting point which emerges
from our analysis is the possibility of an intermediate phase, the
Aharonov-Bohm insulating phase, separating the Mott insulator from the
superfluid.

\subsection{Zero magnetic \& electric frustration}
\label{diagram0}

A first	 estimate for the location of the phase boundary can be
obtained by means of the mean-field approach described in \sect{scmf}.
The results coincide with the first-order perturbative calculation
introduced in \sect{pert} and with the Gutzwiller variational approach
of \sect{Gutz}. In absence of frustration the $\vett{K} = \vett{0}$
mode corresponds to the maximum eigenvalue of the matrix $\mathcal{P}$
($\pi_{max} = \sqrt{18}$) and the transition point is given by
\begin{equation}
         t_{cr} =
         \frac{1}{6 \, \sqrt{2}} \, \sqrt{\trasf{U}_{A,A} (\vett{0})
   \, \trasf{U}_{B,B}(\vett{0})}
\label{tcr00}
\end{equation}

\begin{figure}[hbtp]
\centerline{  \includegraphics[width=8cm]{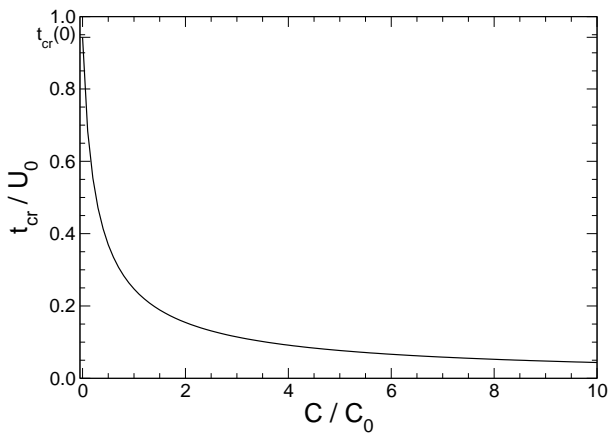}}
       \caption{
   Josephson arrays: dependence of the critical point at
       $f=0$ on the range of the Coulomb repulsion determined by the
       ratio $C/C_0$.}
\label{trans0}
\end{figure}

\begin{figure}[hbtp]
\centerline{  \includegraphics[width=8cm]{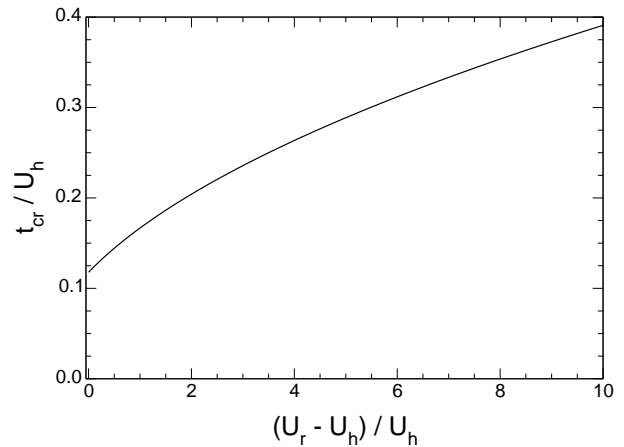}}
       \caption{
   Optical lattices: Dependence of the transition point on
       the difference repulsion in the hubs and the rims}
\label{trans0o}
\end{figure}

In the limit of on-site uniform ($U_r = U_h = 8 U_0$) the SI transition
occurs at the value $ t_{cr}/U_0 = 2 \sqrt{2}\, /3 \approx 0.943$ very
close to the mean field value for a square lattice $t_{cr}/U_0 = 1$ (in
both lattices the average value of nearest neighbours is 4). In the
case of a Josephson array the transition point depends on the range of
the interaction, see Eq.(\ref{Cij}). In the (more realistic) case of a
finite junction capacitance an analytic form is not available and the
numerical phase boundary is shown in \fig{trans0} as a function of the
ratio $C/C_0$.
In the case of optical lattices, see Eq.(\ref{onsite}), the
repulsion is on-site. There is still a weak dependence of the transition
on the difference $U_r-U_h$. As it is shown in \fig{trans0o}, this
dependence is not particularly interesting and in the Monte Carlo
simulation we will ignore it.

As already mentioned, a characteristic feature that emerges in $\TT$
lattices, even in the absence of magnetic frustration, is that the
superfluid order parameter is not homogeneous. This can be already seen
from the eigenvector corresponding to the solution of Eq.(\ref{tcr00}).
Near the transition point the ratio between the order parameter value
on hubs and rims is constant and is related to the ratio of the 
on-site repulsions $\abs{\psi_h/\psi_r} \simeq \sqrt{z_h U_r / z_r 
U_h}$. Phase localization is more robust on hubs ($z_h = 6$) than on 
rims ($z_r = 3$)	because of the larger number of nearest neighbours. 
In order to better understand the modulation of the order parameter we 
analyzed the properties of the superconducting phase using the 
variational approach exposed in \sect{Gutz} (which allows us to study 
the behaviour of $\psi$ also far from the transition).
\begin{figure}
  \centerline{  \includegraphics[width=8cm]{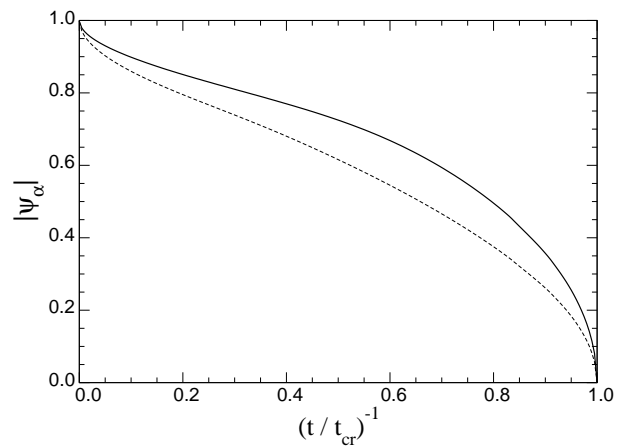}}
       \caption{
   Modulation of the order parameter for zero frustration,
         $\psi_{hub}$ (straight) is always higher than $\psi_{rim}$
         (dashed). The curves are obtained by means of the Gutzwiller
         variational approach.}
\label{figmodul}
\end{figure}
As it can be clearly seen from Fig.\ref{figmodul}, quantum fluctuations
have a stronger effect on the rims than hubs due to the different
coordination number of the two sublattices. Note that this is a pure
quantum mechanical effect, in the classical regime all phases are well
defined	 and $\psi_{hub}=\psi_{rim}= 1$. The transition point (as it
was implicit in the previous discussion) is the same for both
sublattices: there is no possibility to establish phase coherence
between rims if the hub-network was already disordered (and viceversa).

\subsection{Electric frustration}
\label{electric}

\begin{figure}
  \includegraphics[width=4cm]{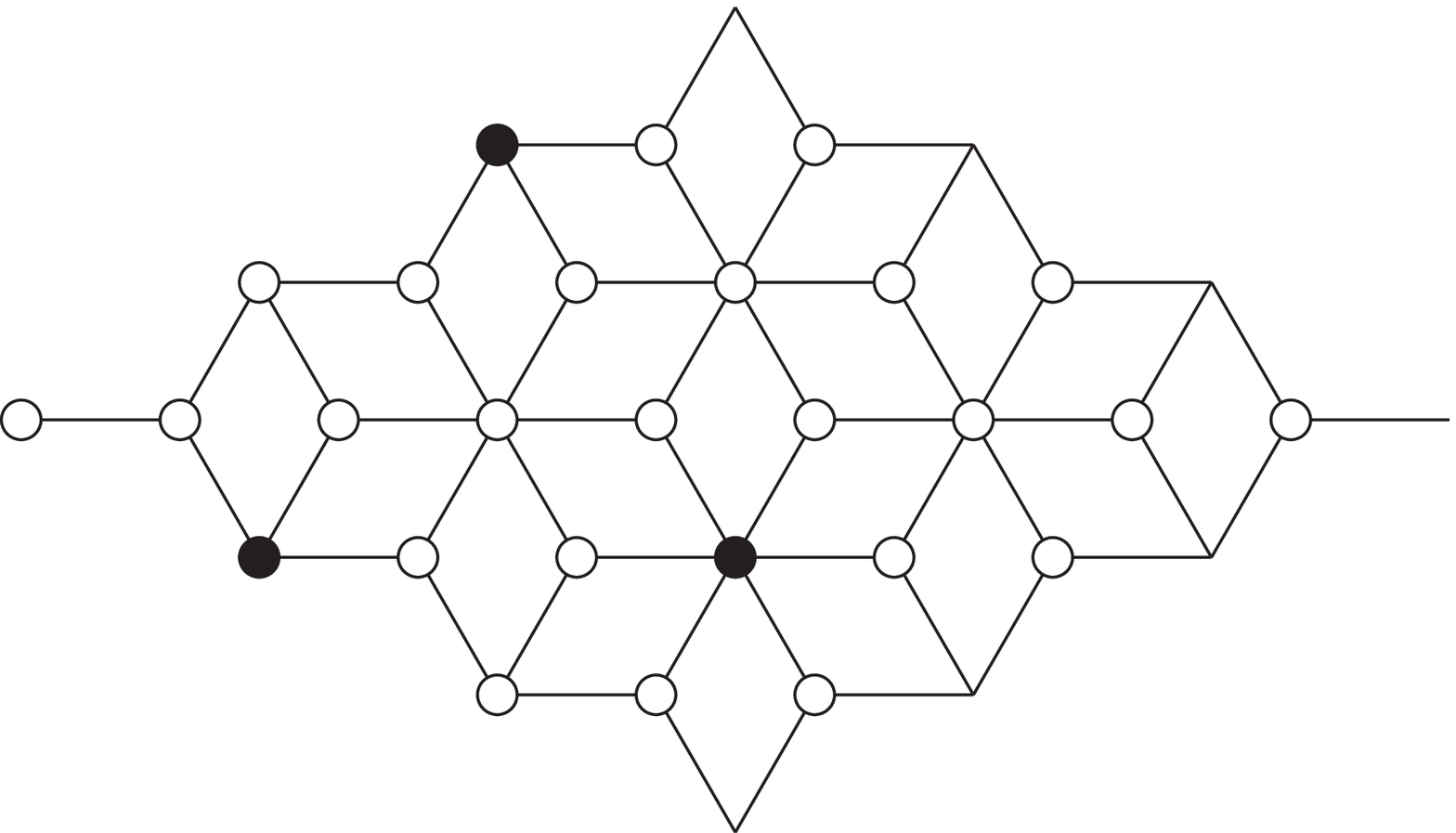} \hspace{0.4cm}
  \includegraphics[width=4cm]{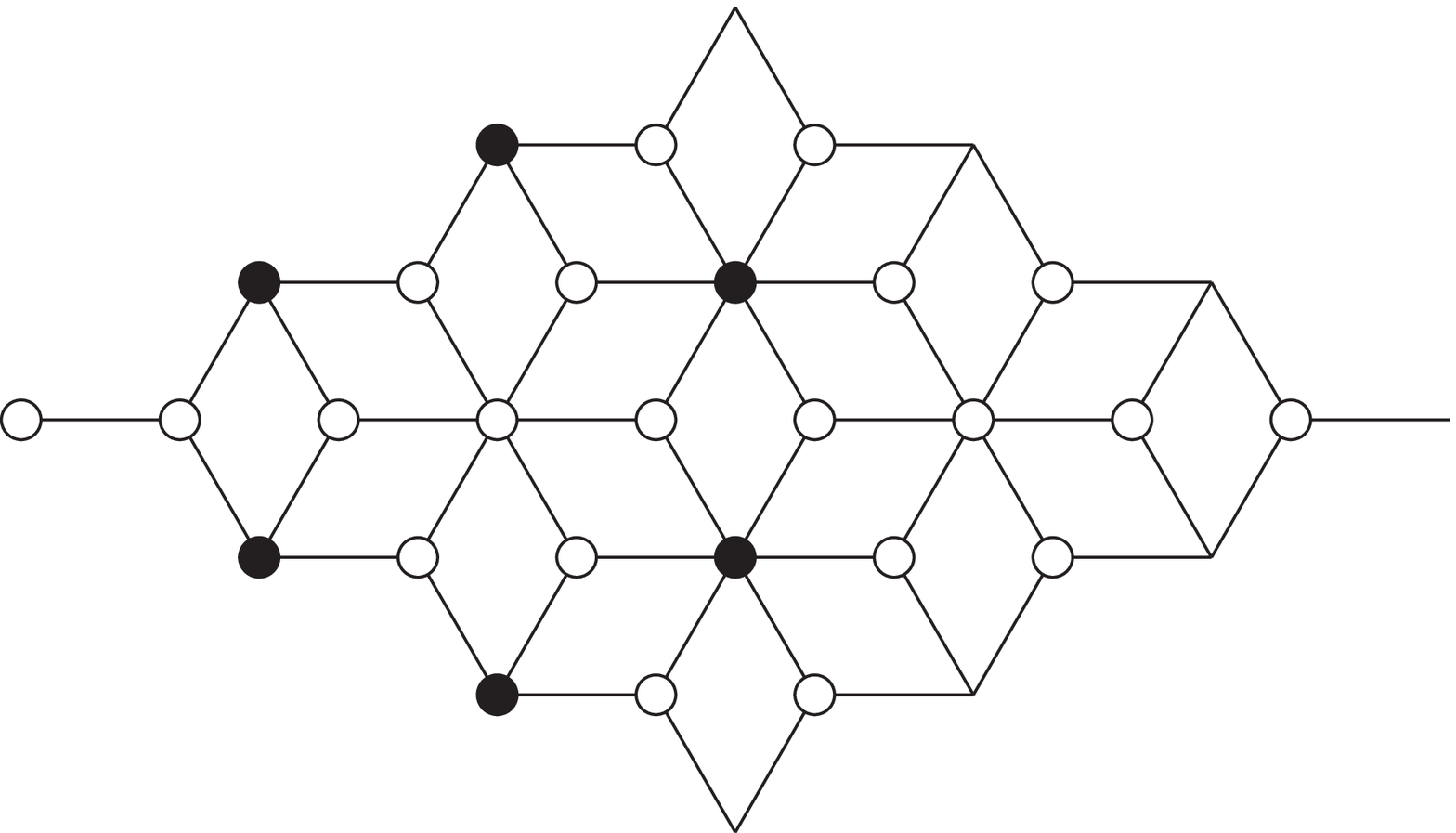} \\
  \vspace{0.4cm}
  \includegraphics[width=4cm]{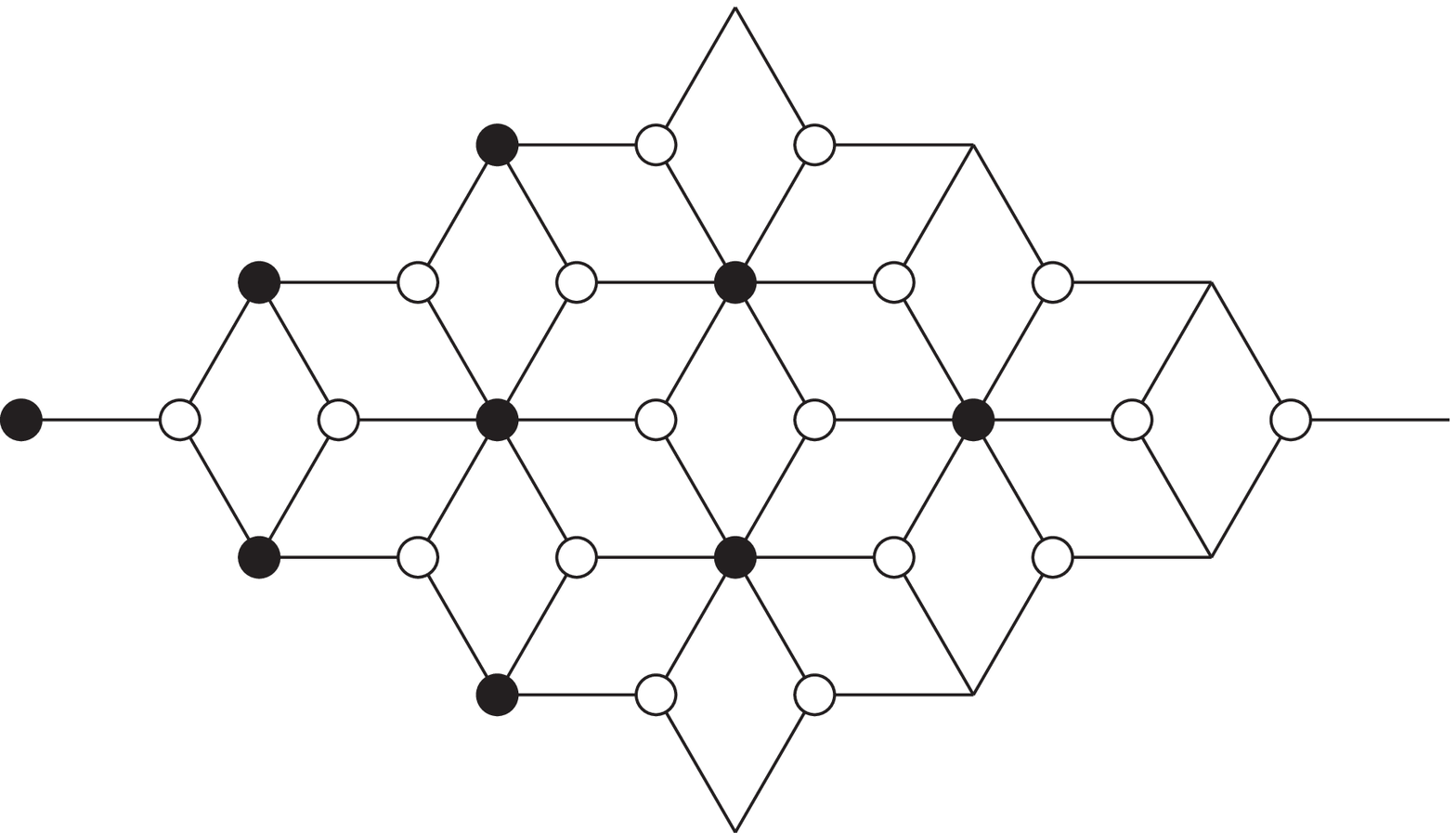} \hspace{0.4cm}
  \includegraphics[width=4cm]{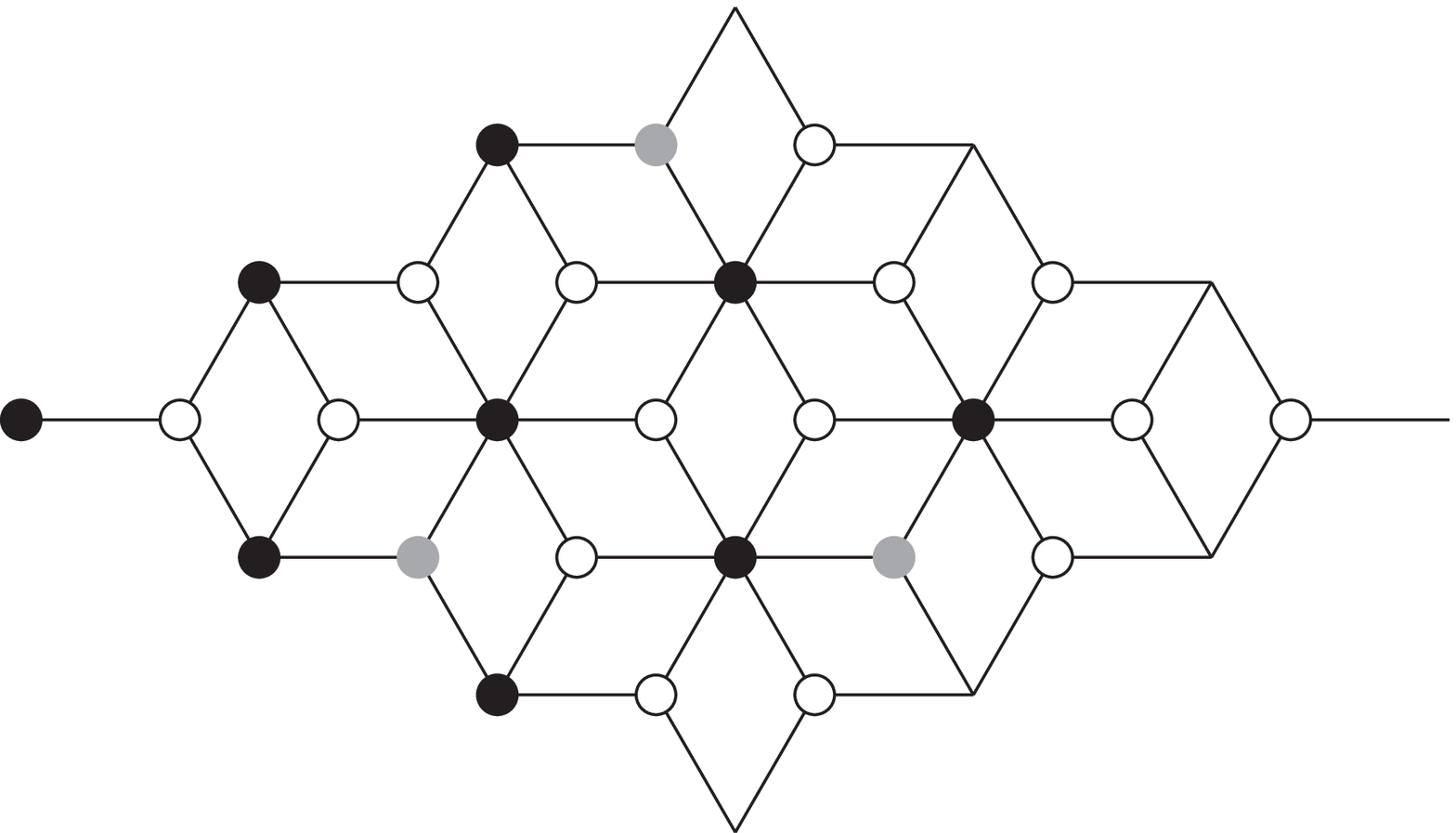}
\caption{
   Ground state configurations of the charges (i.e. at $t=0$)
   for filling $1/9, 2/9,1/3, 4/9$. The different ground states
   occurs on increasing the value of the external charge $n_0$
   The black circles denote those sites that are
   occupied by one Cooper pair. The ground state configurations
   are responsible for the behaviour of the phase correlator and
   hence of the lobe-like structure, Fig.\ref{lobes}, of the
   phase diagram.}
\label{fig:lobeground}
\end{figure}

When an external uniform charge frustration is present, the array
cannot minimize the energy on each site separately, hence frustration
arises. The behaviour of the transition point as a function of the
offset charge shows a typical lobe-structure~\cite{fisher89,bruder93}.
At the mean-field level all the information to obtain the dependence of
the phase boundary on the chemical potential (gate potential for
Josephson arrays) is contained in the zero-frequency transform of the
Green functions $G$ in Eq.(\ref{eq:ejcrit}). The calculation of the
phase-phase correlators, defined by Eq.(\ref{corrG}), is determined, at $T =
0$, once the ground and the first excited states of $\mathcal{H}_U$ is
known. As all the observables are periodic of period one in the offset
charge $n_{0}$ and are symmetric around $n_{0} = 0$, the analysis can
be restricted to the interval $[0,1/2]$. {\em Ground state} charge
configuration in the case of some values of the electric frustration are
shown in Fig.\ref{fig:lobeground}

\begin{figure*}
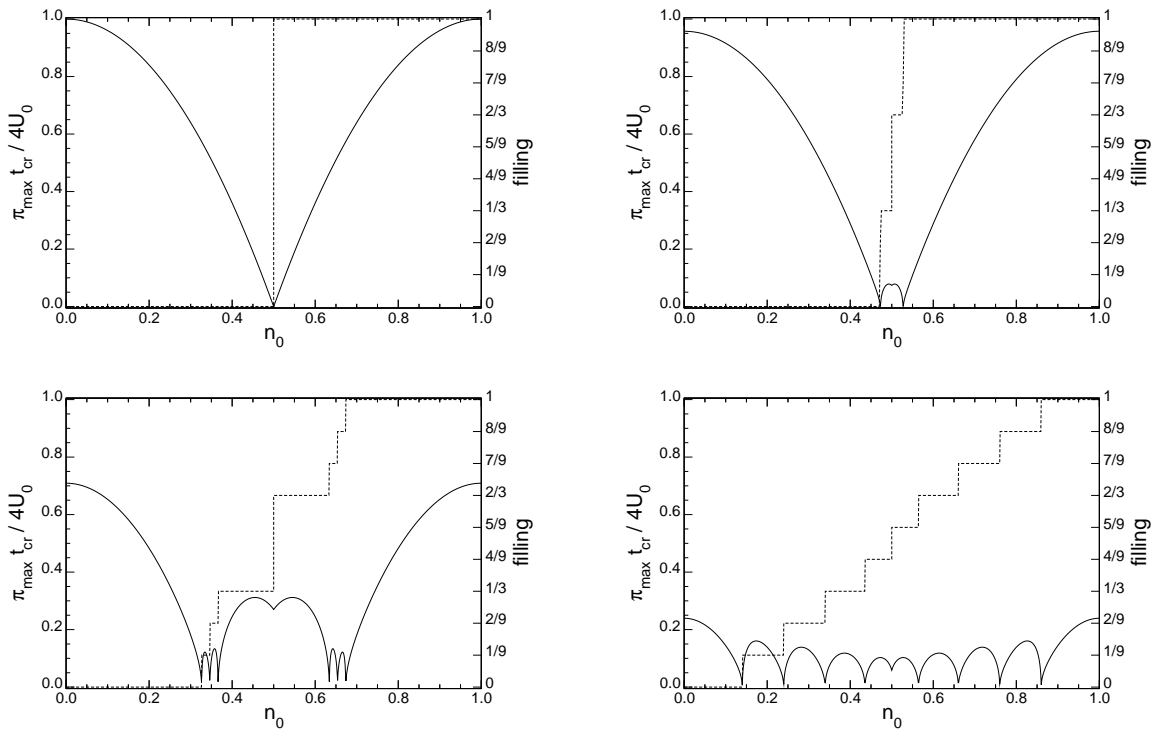

  \label{sections}
  \begin{center}
    \includegraphics[width=7cm]{lobes_def_C000}
    \hspace{1cm}
    \includegraphics[width=7cm]{lobes_def_C001}\\
    \vspace{5mm}
    \includegraphics[width=7cm]{lobes_def_C010}
    \hspace{1cm}
    \includegraphics[width=7cm]{lobes_def_C100}
    \caption{Lobe structures at different values of the
      capacity, i.e. electrostatic range (respectively $C = 0,
      10^{-2}, 10^{-1}, 1$).
      The dashed lines point out the discrete filling of the ground
      state. Pictures on the right are magnifications of the
      highlighted areas in the left ones.}
    \label{lobes}
  \end{center}
\end{figure*}

The phase diagram in the  presence of charge frustration has a lobe
structure~\cite{fisher89} in which, progressively on increasing the
external charge, the filling factor increases as well. In the case of
finite range charging interaction also Mott lobes with fractional
fillings appear~\cite{bruder93}. An analytical determination of the
ground state of the charging Hamiltonian for generic values of the
external charge is not available. We considered rational fillings of
the whole lattice as made up of periodic repetitions of a partially
filled super-cell of size comparable with the range of the interaction
$\mathcal{U}_{i,j}$ and then constructed a Wigner crystal for the
Cooper pairs with this periodicity. For $C/C_0 \leq 1$ a $3 \times 3$
super-cell turns out to be sufficient. Given a certain rational
filling $p/q$, the corresponding charging energy is given by
\begin{equation*}
         E_{ \{n_i\}}(\frac{p}{q}, n_{0}) = 3 N \, \frac{e^2}{C_0} \, \left(
         n_{0}^2 - 2 \frac{p}{q} n_{0} + \frac{C_0}{N} \, \sum_{i,j} n_i
         \mathcal{C}^{-1}_{i,j} n_j \right)
\end{equation*}
where $N$ is the number of cells in the system and $\{n_i\}$ is the
particular realization of the filling.

This defines a set of parabolas which allow to determine the
sequence of ground states. The variation of the ground state
configurations as a function of gate charge gives to the phase boundary
a characteristic structure made of lobes, as shown in \fig{lobes}. The
longer is this range of the electrostatic interaction the richer is the
lobe structure.

As can be seen in \fig{lobes} when the interaction is purely on-site
there is only one lobe that closes at half filling when the degeneracy
between the empty ground state and the extra-charged one leads to
superconductivity for arbitrarily small $t$.	 As soon
as the range becomes finite, other fillings come into play. An
interesting feature typical of the $\TT$ lattice is that at $n_0=1/2$
the half filled state is  not		the ground state (see \fig{lobes}).

Finally, we recall that the presence of the offset breaks the
particle-hole symmetry and thus the universality class of the phase
transition change~\cite{fisher89}. This can be seen from the expansion at small
$\omega$ of the correlator (\equ{Gtrasf}) that enters the quadratic
term of the Wilson-Ginzburg-Landau functional. With $n_{0}$ also terms
linear in $\omega$ enter the expansion and the dynamical exponent $z$
changes from $1$ to $2$.

\subsection{Magnetic frustration and Aharonov-Bohm insulating phase}
\label{magnetic}

\begin{figure}[htbp]
   \centerline{\includegraphics[width=8cm]{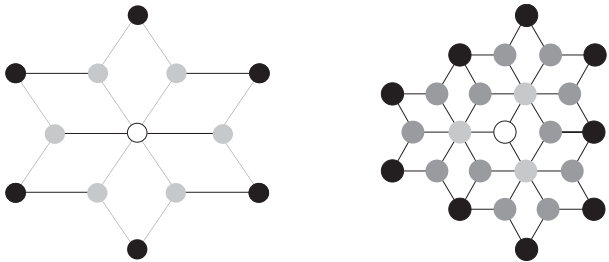}}
   \caption{
     Aharonov-Bohm cages. Particles that starts on white sites
     can't go further than black sites, due to destructive interference.
     In fact, $f = 1 / 2$ means $\pi$ phase shift around a plaquette.
     In square lattices this could not happen because of the escape
     opportunity given by straight lines.}
   \label{cages}
\end{figure}

The outgrowing interest in $\TT$ lattices is especially due to their
behaviour in the presence of an externally applied magnetic field. The
presence of a magnetic field defines a new length scale, the magnetic
length. The competition between this length and the lattice
periodicity generates interesting phenomena such as the rising of a
fractal spectrum \`a la Hofstadter. In $\TT$ lattices perhaps the most
striking feature is the complete localization in a fully frustrating field
($f= 1 / 2$). This is due to destructive interference along all paths that
particles could walk on, when the phase shift around a rhombic
plaquette is $\pi$ (see \fig{cages}). Is there any signature of this
localization (originally predicted for tight-binding models) in the
quantum phases transition between the Mott and the superconducting
phases? This is what we want to investigate in this section.

\begin{figure}[htbp]
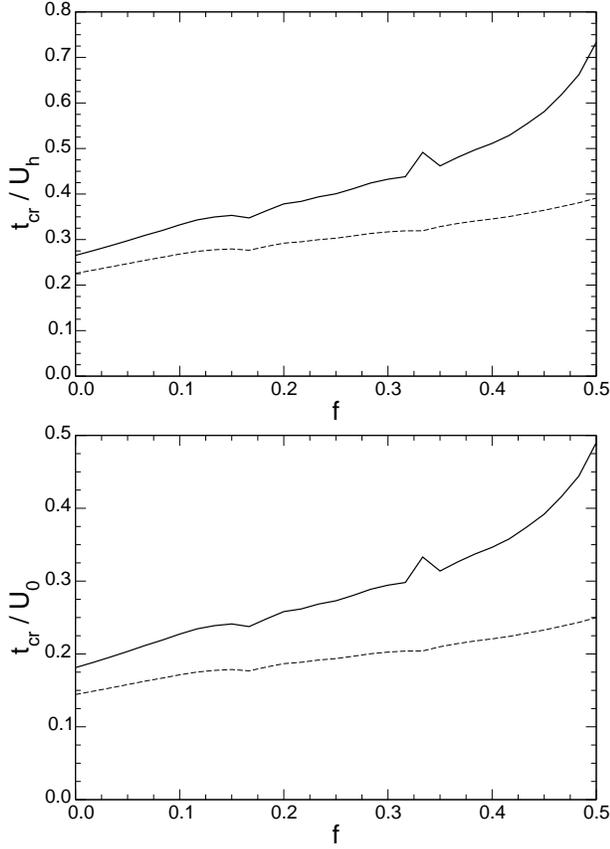

  \includegraphics[width=8cm]{perturb_finrange}
  \includegraphics[width=8cm]{perturb_norange}
  \caption{Phase boundary in presence of a magnetic field in $\TT$:
    straight line is the perturbative result, mean field is dashed. Upper:
    JJAs with $C/C_0 = 1$; lower: optical lattices with $U_r - U_h
    = 0.5 U_h$. Note the highly pronounced peak at $f = 1 / 2$ in
    contrast to the square lattice case.}
  \label{figmag}
\end{figure}

In order to determine the phase boundary at $T = 0$ we can follow
either the mean field approach of \sect{scmf} or the perturbative
theory presented in \sect{pert}. We remind that while the first
approach signals the disappearance of the superfluid phase, the
perturbation expansion indicates where the Mott phase ends. The
results of both approaches are shown in \fig{figmag}. Commensurate
effects are visible in the phase boundary of \fig{figmag} at rational
fractions $f = p / q$ of the frustration. The results presented are
quite generic. We decided to show, as a representative example, the
results for a JJ array with capacitance ratio $C/C_0 = 1$ and an
optical lattice with $U_r - U_h = 0.5 U_h$. The peak at $f = 1 / 2$,
characteristic of the $\TT$ lattice is due to the presence of the
Aharonov-Bohm cages.

Although there is a difference between the mean-field and the strong
coupling calculation, they both confirm the same behaviour. A very
interesting point however emerges at half-filling. It is worth to
stress again  that while the mean-field shows the
disappearance of the superconducting phase, the strong coupling
expansion indicates where the Mott gap vanishes and hence charges can
condense. The vanishing of the gap can be associated to boson
condensation {\em only} if bosons are
delocalized. This is the case for the whole range of frustrations
except at $f=1/2$.  In the fully frustrated case the excitation gap
{\em vanishes} but the excited state (the extra boson on a hub) still
remains localized due to the existence of the Aharonov-Bohm cages. This
may lead to the conclusion that at fully frustration there is an
intermediate phase where the system is {\em compressible} (the Mott gap
has been reduced to zero) with {\em zero superfluid density} (the
bosons are localized in the Aharonov-Bohm cages).

At this level of approximation there is no way to explore
further this scenario. In order to assess
the existence of the intermediate phase a more accurate location of the
phase boundaries is necessary. We will discuss the possible existence
of the {\em Aharonov-Bohm insulator} by means of Monte Carlo
simulations in the next section.

\subsection{MonteCarlo methods}
\label{QMC methods}

The simulations are performed on an effective classical model obtained
after mapping the model of Eq.(\ref{hamiltonian}) onto a $(2+1)$ XY
model. Our main interest in performing the Monte Carlo simulation is to
look for signatures of the Aharonov-Bohm insulator. As its existence
should not depend on the exact form of the repulsion
$\mathcal{U}_{i,j}$ we chose the simplest possible case in which the
repulsion is on-site and $U_h \sim U_r$. The details of the mapping are
described in  Refs.~\onlinecite{jacobs88,wallin94} and are briefly
reviewed in \sect{3dxymap}. The effective action $\mathcal{S}$ (at zero
charge frustration) describing the equivalent classical model is
\begin{eqnarray}
\label{isotropic}
    \nonumber
    \mathcal{S} & = &
    K \,
    \sum_{
    \begin{array}{c}
        \langle i,\, j \rangle , k
    \end{array}
    }
    \left[ 1 - \cos \left(\varphi_{i,k} -
    \varphi_{j,k} -
    A_{i,j}\right) \right] \\
     & + &
    K \,
    \sum_{
    \begin{array}{c}
        i, \langle k,\, k' \rangle
    \end{array}
    }
    \left( 1 - \cos \left(\varphi_{i,k} -
    \varphi_{i,k'} \right) \right) \;.
\end{eqnarray}
where the coupling $K$ is $\sqrt{t/U}$. The index $k$ labels the extra
(imaginary time) direction which takes into account the quantum
fluctuations. The simulations where performed on $L \times L \times
L_{\tau}$ lattice with periodic boundary conditions. The two
correlation lengths (along the space and time directions) are related
by the dynamical exponent $\mbox{z}$ through the relation $\xi_{\tau}
\sim \xi^{z}$. For zero magnetic frustration, because of the
particle-hole symmetry (we consider only the case $n_0=0$) holds
$z=1$. As we will see this seems not to be the case at fully
frustration because of the presence of the Aharonov-Bohm cages.

The evaluation of the various quantities have been obtained averaging
up to $3 \times 10^{5}$ Monte Carlo configurations for each one of the $10^2$
initial conditions, by using a standard Metropolis algorithm. Typically
the first $10^{5}$ were used for thermalization. The largest lattice
studied was $24\times 16 \times 24$ at fully frustration and $48 \times
48 \times 48$ at $f=0$. This difference is due to the much larger
statistics which is needed to obtain sufficiently reliable data. While
in the unfrustrated case we took a cube of length $L$ in the fully
frustrated case it turned out to be more convenient to consider (but
will discuss other lattice shapes) an aspect ratio of $2/3$. With this
choice the equilibration was simpler probably due to a different
proliferation of domain walls\cite{korshunov02,cataudella03}.

In order to characterize the phase diagram we studied the superfluid
stiffness and the compressibility of the Bose-Hubbard model on a $\TT$
lattice. The compressibility, $\kappa$, is defined by $ \kappa =
\partial^2 \mathcal{F} / \partial V_0^2 $ where $\mathcal{F}$
is the free energy of the system and $V_0$ the chemical potential for
the bosons. By employing  the Josephson relation in imaginary time,
see Ref.\cite{wallin94}, the compressibility can be expressed as the
response of the system to a twist in imaginary time, $\varphi_{i,k}
\rightarrow \varphi_{i,k} + \theta _{\tau}\,k$, i.e.
\begin{equation}
         \kappa  = \left.
      \frac{\partial^2	 \mathcal{F}(\theta _{\tau})}{ \partial{\theta^2 _{\tau}}}
      \right|_{\theta _{\tau}=0} \ \; .
\label{compressdef}
\end{equation}

The superfluid stiffness is associated to the free energy cost to impose a
phase twist in a direction $\vett{e}$, i.e.
$
      \varphi_{i} \rightarrow \varphi_{i} + \theta_{\vett{e}} \, \vett{e}
      \cdot \vett{r}_i
$,
through the array
\begin{equation}
      \gamma  = \left.
      \frac{\partial^2	 \mathcal{F}(\theta _{\vett{e}})}{
      \partial{\theta^2_{\vett{e}}}}
      \right|_{\theta _{\vett{e}}=0}  \;\; .
\label{stiffdef}
\end{equation}

\subsubsection{f=0}

In the case of unfrustrated system we expect that the transition
belongs to the $3D-XY$ universality class. Close to the quantum
critical point $\kappa \sim \xi^{-1}$, the corresponding finite size
scaling expression for the compressibility reads
\begin{equation}
\label{finsizesccom}
      \kappa = L^{-(d-z)} \, \widetilde{\kappa} \left(
      L^{1 / \nu} \, \frac{K - K_{c}}{K_c}, \, \frac{L_{\tau}}{L^z} \right)
\end{equation}
\begin{figure}
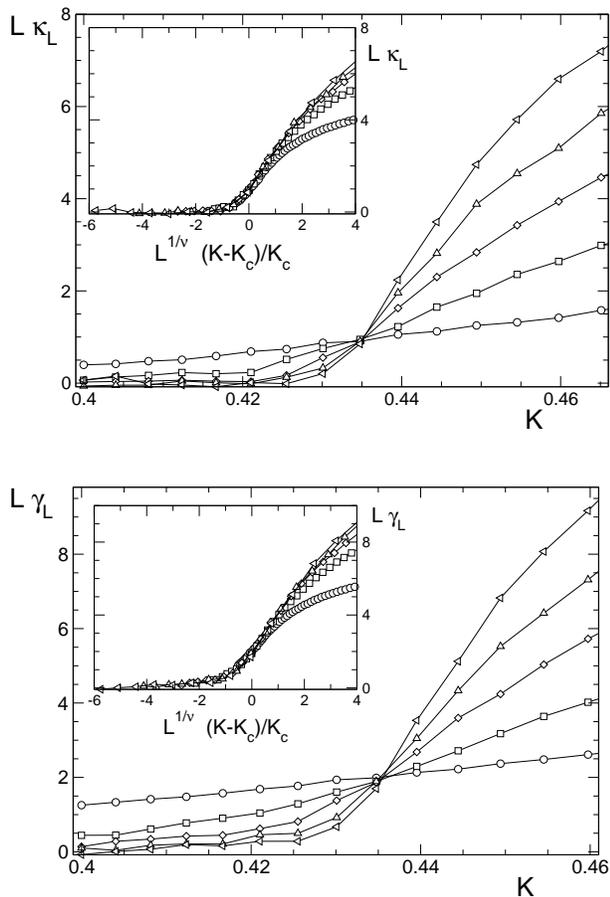

  \includegraphics[width=8cm]{compress00}\\
  \vspace{7mm}
  \includegraphics[width=8cm]{stiffness00}
  \caption{$f=0$:
    a)Scaling (main) and data collapse (inset) of the compressibility
    for the unfrustrated case.
    b)The same as in a) for the superfluid stiffness.
    All the systems have aspect ratio $L_{\tau} = L_y = L_x$ with
    $L_{\tau} = 6$ (circles), 12 (squares), 18 (diamonds), 24
    (triangles	 up), 30 (triangles down)}
  \label{scalingf0}
  \label{collapsef0}
\end{figure}

An analogous expression holds for the finite size-scaling behaviour
of the stiffness
\begin{equation}
  \label{finsizescstif}
  \gamma = L^{-(d+z-2)} \, \widetilde{\gamma} \left(
  L^{1 / \nu} \, \frac{K - K_{c}}{K_c}, \, \frac{L_{\tau}}{L^z} \right)
\end{equation}
The expected exponent is $\nu = 2/3$ as it is known from the properties
of the three-dimensional $XY$ model.

The results of the simulations for the compressibility and for the
stiffness are reported in Fig.\ref{scalingf0}. Finite size scaling
shows that the SI transition occurs at
\begin{equation}
  K_c = 0.435 \pm 0.0025 \;\; .
\end{equation}
As expected the unfrustrated case follows remarkably well the standard
picture of the Superfluid-Mott Insulator quantum phase transition. In
the absence of the magnetic field the system defined by
Eq.(\ref{isotropic}) is isotropic in space-time and therefore the
stiffness and the compressibility have the same scaling and critical
point.

\subsubsection{$f=\frac{1}{2}$}

The situation changes dramatically in the fully frustrated system. In
this case an anisotropy in space and time directions arises because of
the presence of the applied magnetic field which frustrates the bonds
in the space directions (see the r.h.s of Eq.(\ref{isotropic})). This
field induced anisotropy is responsible for the different behaviour of
the system to a twist in the time (compressibility) or space
(stiffness) components.

As already observed in the classical case~\cite{cataudella03}, the
Monte Carlo dynamics of frustrated $\TT$ systems becomes very slow.
This seems to be associated to the presence of zero-energy domain walls
first discussed by Korshunov in Ref.~\onlinecite{korshunov02}. This
issue is particulary delicate for the superfluid stiffness. In this
case the longest simulations had to be performed. Moreover in order to
alleviate this problem we always started the run deep in the superfluid
state and progressively increased the value of the Hubbard repulsion
$U$. Also the choice of the lattice dimensions turned out to be
important. We made the simulations on $12 \times 8 \times 12$, $18
\times 12 \times 18$, and $24 \times 16 \times 24$ systems and found
out that by choosing this aspect ratio along the $x$ and $y$ directions
thermalization was considerably improved.

\begin{figure}
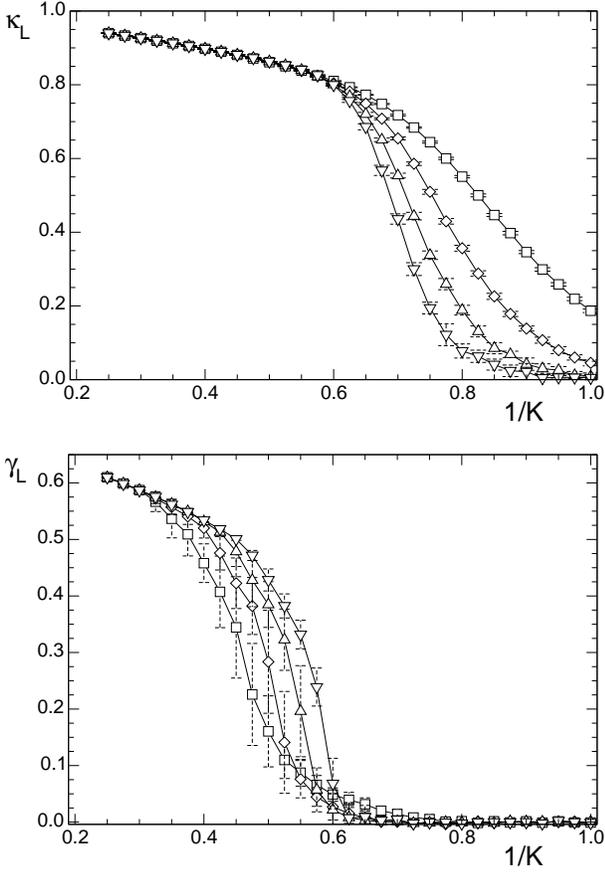

  \centerline{\includegraphics[width=8cm]{compress05}}
  \hspace{0.2cm}
  \centerline{\includegraphics[width=8cm]{stiffness05}}
  \caption{
    $f=1/2$:Compressibility (upper panel) and stiffness (lower panel)
    assuming the aspect ratio $L \times L \times 2L/3$.
    Different symbols corresponds to $L =$ 12 (circles), 18 (squares), 24
    (triangles up), and 30 (triangles down).}
  \label{scalinone05}
\end{figure}

The results of the simulations are reported in Fig.\ref{scalinone05}
for the compressibility and for the stiffness. As it appears from the
raw data of the figure it seems that the points at which the
compressibility and the stiffness go to zero are different. An
appropriate way to extract the  critical point(s) should be by
means of finite size scaling.

\begin{figure}
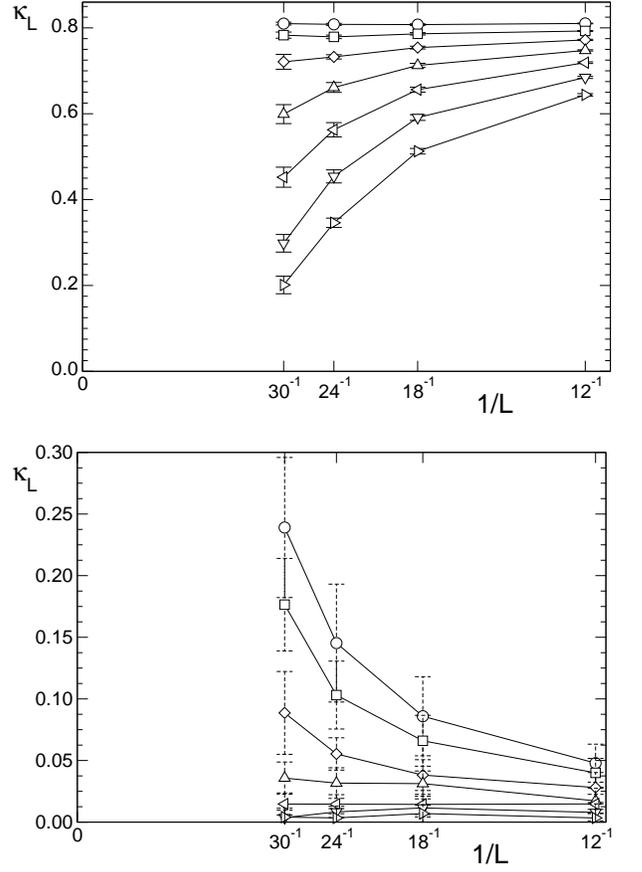

  \centerline{\includegraphics[width=8cm]{comprscal05}}
  \hspace{0.2cm}
  \centerline{\includegraphics[width=8cm]{stiffscal05}}
  \caption{$f=1/2$: Compressibility (upper panel) and the stiffness
    (lower panel) as a function of the $L_t$ size of the system for different
    values of $K$.
    Data corresponds to $1/K = 0.6$(circles), $0.625$(squares),
    $0.65$(diamonds), $0.675$(triangles up), $0.7$(tr. left),
    $0.725$(tr. down), $0.75$(tr. right).}.
  \label{L05}
\end{figure}

As a first attempt we assumed that the transition is in the same
universality class as for the unfrustrated case and we scaled the data as in
Fig.\ref{scalingf0}. Although the scaling hinted at the existence of two
different critical points for the Mott to Aharonov-Bohm insulator and for the
Aharonov-Bohm insulator to superfluid transitions respectively, the
quality of the scaling points was poor.
In our opinion this observation may suggest that the
scaling exponents for the fully frustrated case are different as the
one for the \textbf{direct} Mott Insulator to Superfluid phase
transition at $f\neq 1/2$. In order to extract more tight
bounds on the existence of this phase we analyzed the size dependence
of the observables without any explicit hypothesis on the scaling
exponent (which we actually do not know). The results are presented in
Fig.\ref{L05}.
The data of Fig.\ref{L05} seem to indicate that there is a window
$$ 0.65 \le K^{-1} \le 0.7 $$
where the system is compressible but not superfluid! This new phase,
the Aharonov-Bohm insulator, is the result of the subtle
interplay of the $\TT$ lattice structure and the frustration
induced by the external magnetic field. Our simulations cannot firmly
determine the existence of two separate critical points	since
we were not able to improve their accuracy and study larger lattices.
However we think that, by combining both the analytical results and the
Monte Carlo data we have a possible scenario for the phase diagram of the
frustrated BH model on a $\TT$ lattice.

\begin{figure}
  \includegraphics[width=8cm]{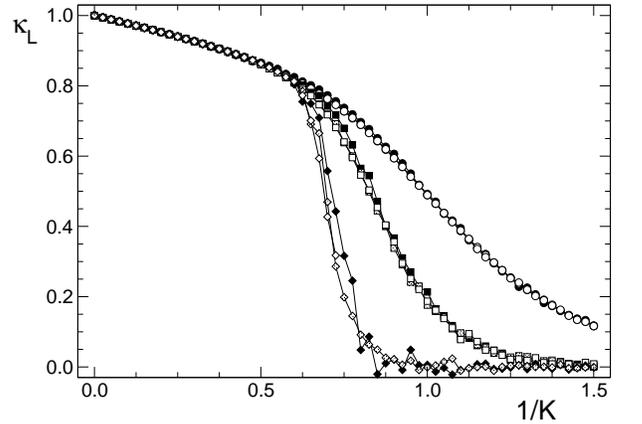}
  \caption{
    Compressibility as a function of $K=\sqrt{t/U}$ for
    different values of the system sizes. Different symbols
    corresponds to different lengths $L_\tau$ in the time dimension:
    6 (circles), 12 (squares), and 30 (diamonds). Different fillings are
    different spatial sizes $L_x \times L_y$: $6 \times 4$ (black),
    $12 \times 8$ (gray), and $18 \times 12$ (white).
    The compressibility depends strongly on $L_{\tau}$ but very
    weakly on $L_x \times L_y$. }
  \label{ABphase}
\end{figure}

Further evidence of the existence of the AB cages can be obtained by
analyzing the anisotropy in space and time directions of the phase
correlations. For this purpose we considered the compressibility as a
function of $L$ and $L_{\tau}$ separately. The idea is that because of
the AB cages the correlations are short-ranged in the space directions
(bosons are localized) while there are longer ranged correlations in
the time direction. Indeed the dependence of the compressibility on the
system dimensions is strong when one changes $L_{\tau}$ while it is
rather weak when the space dimensions are varied as shown in
Fig.\ref{ABphase}. This hints at the fact that the Aharonov-Bohm phase
is a phase in which the gap has been suppressed (correlation in the
time dimension) but where the bosons are localized (short-range
correlations in space).

The Monte Carlo simulations just discussed provide evidence for the
existence of a new phase between the Mott insulator and superfluid. Due
to the finite size of the system considered and to the (present) lack
of a scaling theory of the two transitions, we cannot rule out other
possible interpretations of the observed behaviour of the Monte Carlo
data. A possible scenario which is compatible with the simulations (but
not with the result of the perturbation expansion~\cite{footnote3}) is
that a single thermodynamic transition is present in the $2+1$
dimensional system but the phase coherence is established in a two step
process.	 First the
system becomes (quasi) ordered along the time direction, then, upon
increasing the hopping the residual interaction between these
\virg{quasi-one-dimensional} coherent tubes go into a three-dimensional
coherent state driven by the residual coupling between the tubes. In
more physical terms the \virg{tubes} represent the boson localized in
the AB cages and the residual hopping is responsible for the transition
to the superfluid state. This means that the intermediate state that we
observe is due to a one- to three-dimensional crossover that takes
place at intermediate couplings.

\section{Conclusions}
\label{discuss}

In this work we exploited several methods, both analytic and numerical,
in order to determine the phase diagram of a Bose-Hubbard model on a
$\TT$ lattice. Differently from previous studies on $\TT$ networks we
analyzed the situation where the repulsion between bosons (or Cooper
pairs for Josephson arrays) becomes comparable with the tunnelling
amplitude (Josephson coupling in JJAs) leading to a quantum phase
transition in the phase diagram. Up to now the attention on
experimental implementations has been confined to Josephson networks.
As discussed in \sect{optical}, the $\TT$ lattice can also be realized
in		optical lattices. The possibility to experimentally study
frustrated $\TT$ optical lattices open the very interesting possibility
to observe subtle interference phenomena associated to Aharonov-Bohm
cages also with cold atoms. Having in mind both the realization in
Josephson and optical arrays, we studied a variety of different
situations determined by the range of the boson repulsion including
both electric and magnetic frustration. Although in the whole paper we
concentrated on the $T=0$ case, in this discussion we will also comment
on the finite temperature phase diagram.

\begin{figure}
  \includegraphics[width=8cm]{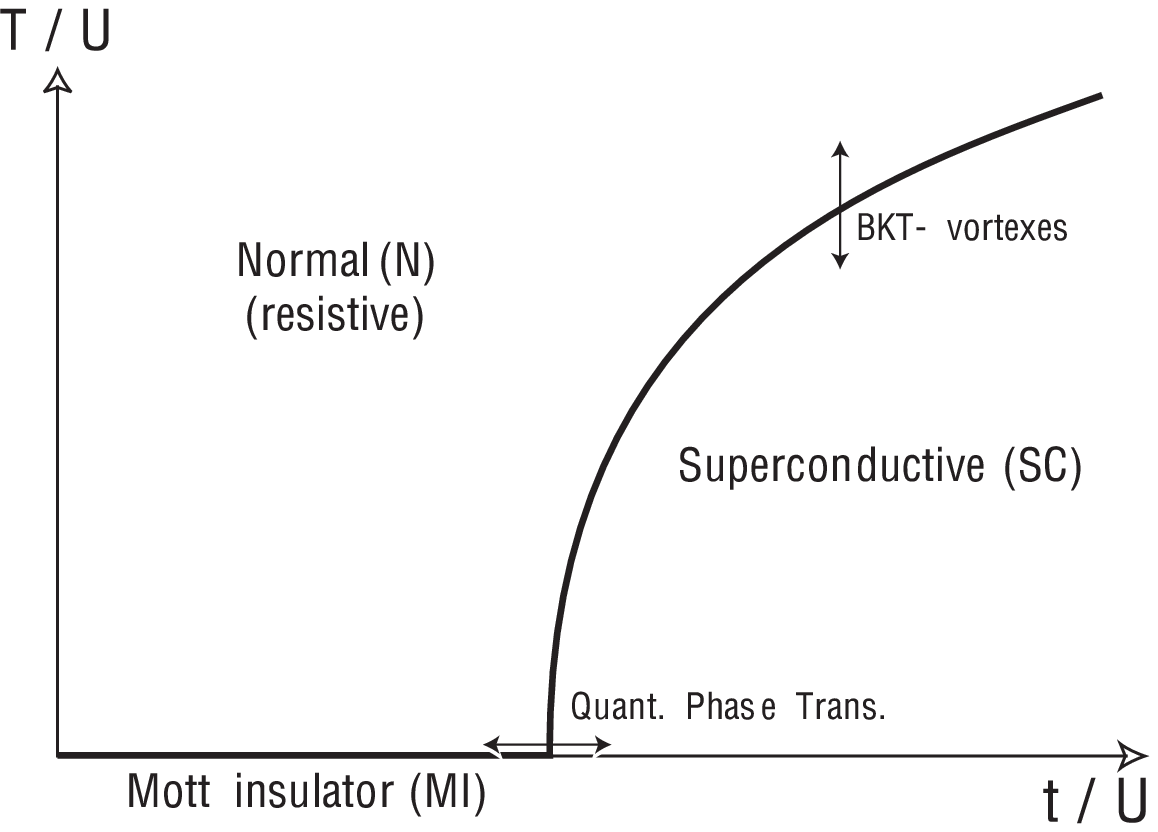}\\
  \vspace{5mm}
  \includegraphics[width=8cm]{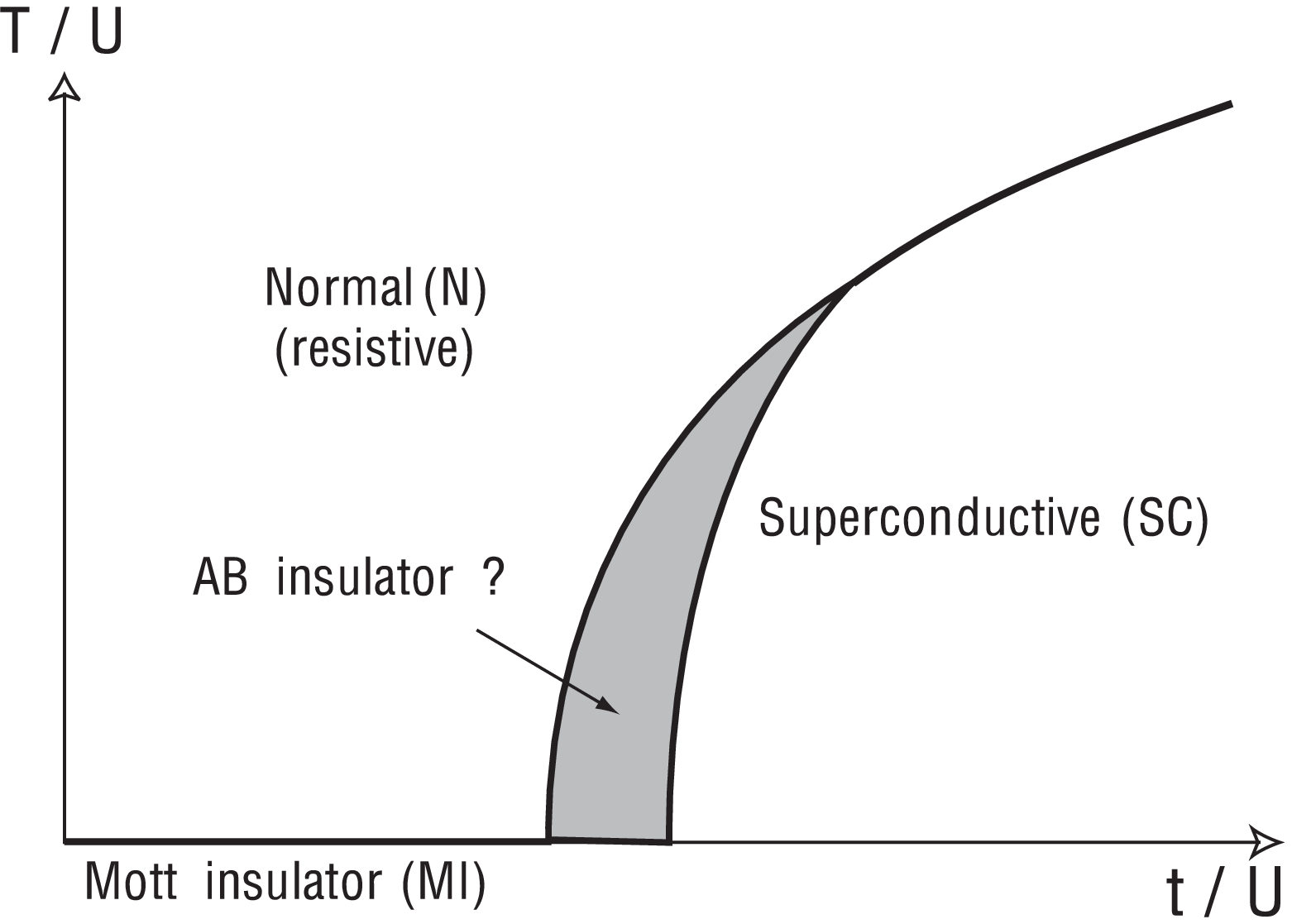}
  \caption{
    A possible phase diagram of an array with $\TT$ lattice.
    In the unfrustrated case (upper panel) we sketch the standard
    phase diagram which leads at $T=0$ to the SI transition. On the
    lower panel we present a possible scenario that emerges from our
    work. An new phase appears separating the normal from the
    superfluid phases.}
  \label{ABphasediagram}
\end{figure}

The peculiarity of the lattice symmetry already emerges for the
unfrustrated case. The superfluid phase is not uniform but it has a
modulation related to the presence of hubs and rims with different
coordination number. As a function of the chemical potential (gate
charge) the transition has a quite rich structure due to the different
boson super-lattices which appear as the ground state.

As a function of the magnetic field the SI transition has the
characteristic butterfly form. In the fully frustrated case, however,
the change is radical and we find indications that the presence of the
Aharonov-Bohm cages can lead to the appearance of a new phase, the
Aharonov-Bohm insulator. This phase should be characterized by a finite
compressibility and zero superfluid stiffness. A sketch of the phase
diagram is shown in Fig.\ref{ABphasediagram}. With the help of Monte
Carlo simulations we were able to bound the range of existence of the
new phase. Unfortunately we have to admit that our results are not
conclusive and, as discussed in the previous section, an alternative
scenario is also possible. Nevertheless, we think that the existence of an
intermediate phase is a very appealing possibility worth to being
further investigated.

How is it possible to experimentally detect such a phase? In Josephson arrays,
where one typically does transport measurement, the AB-insulator should be
detected by looking at the temperature dependence of the linear
resistance. On approaching the zero temperature limit, the resistance
should grow as $T^{\delta}$ differently from the Mott insulating phase
where it has an exponential activated behaviour. In optical lattices
the different phases can be detected by looking at the different
interference pattern (in the momentum density or in the
fluctuations~\cite{altman04}). A detailed analysis of the experimental
probe will be performed in a subsequent publication.

There are several issues that remain to be investigated. It would be
important, for example, to see how the phase diagram of the frustrated
system (and in particular the Aharonov-Bohm	 phase) is modified by a
finite range of $\mathcal{U}_{i,j}$ and/or the presence of a finite
chemical potential. An interesting possibility left untouched by this
work is to study the fully frustrated array at $n_0=1/2$. In this case
(for on-site interaction) the superfluid phase extends down to
vanishing small hopping. In this case a more extended AB insulating
phase could be more favoured, and thus more clearly visible.

\begin{acknowledgments}
We gratefully acknowledge helpful discussions with D. Bercioux, L. Ioffe and
M. Feigelman.
This work was supported by the EU (IST-SQUBIT, HPRN-CT-2002-00144)
\end{acknowledgments}

\appendix

\section{(2+1)D XY mapping} \label{3dxymap}

We give here some of the technical details of the mapping from the QPM
to a $(2+1)D-XY$ model. The latter one is particularly easy to be
simulated numerically: the state of the system and the effective
action are both expressed in terms of phases on a	 $3D$ lattice. Being
$n$ and $\varphi$ canonically conjugated, it is possible to represent
$n$ as $-\imath \pderiv{}{\varphi}$ and get the so-called quantum
rotor Hamiltonian. For the sake of simplicity we consider a diagonal
capacitance matrix.
\begin{eqnarray}
    \nonumber
    \Ham & = & \mathcal{H}_t + \mathcal{H}_U \\
    \nonumber
    \mathcal{H}_U & = & - \frac{U}{2} \sum_{\vett{r}}
    \npderiv{2}{}{\varphi_\vett{r}} \\
    \mathcal{H}_t & = &
    - t \sum_{\langle i,j \rangle}
    \cos (\varphi_i - \varphi_j -
    A_{i,j})
\end{eqnarray}

The partition function can be rewritten in a more convenient way
using the Trotter approximation:
\begin{eqnarray}
    \nonumber
    \mathcal{Z} & = & \traccia{(e^{- \frac{\beta}{L_{\tau}} (\mathcal{H}_t
    + \mathcal{H}_U)})^{L_{\tau}}}\\
                & = & \lim_{L_{\tau} \rightarrow \infty} \traccia{(e^{-
                \Delta \tau \mathcal{H}_U} e^{- \Delta \tau
                \mathcal{H}_t} + o(\Delta \tau^2) )^{L_{\tau}}}
\end{eqnarray}
where $\hbar \tau$ is imaginary time and $\Delta \tau = \beta /
L_{\tau}$ is the width of a time slice. The limit $\Delta \tau
\rightarrow 0$ must be taken to recover the underlying quantum problem.

Introducing complete sets of	 states
$\ket{\conf{\varphi(\tau_k)}}$ with periodic boundary conditions on
times ($\tau_0 = 0 \ \equiv \ \tau_{L_{\tau}} = \beta$) the trace can
be written as
\begin{equation}
    \mathcal{Z} = \int \mathcal{D}\varphi \prod_{k=0}^{L_{\tau}}
    \matrice{e^{- \Delta \tau \, \mathcal{H}_U} e^{- \Delta \tau \,
    \mathcal{H}_t}}%
    {\conf{\varphi(\tau_{k+1})}}{\conf{\varphi(\tau_{k})}}
\end{equation}
Since the	 states $\ket{\conf{\varphi(\tau_k)}}$ are eigenstates of
$\mathcal{H}_t$, the calculation is reduced to the evaluation
of the matrix elements
\begin{equation}
    \matrice{e^{- \Delta \tau \, \mathcal{T}}}%
    {\conf{\varphi(\tau_{k+1})}}{\conf{\varphi(\tau_{k})}}.
\end{equation}
the matrix elements can be furtherly simplified going back to the charge
representation (or angular momentum, since $n$ is the generator of
$U(1)$ for the $XY$ spin of a site):
\begin{equation}
    \sum_{\conf{J^{\tau}}} \prod_i e^{-
    \frac{U \, \Delta \tau}{2} [J^{\tau}_i]^2} e^{\imath
    \, J^{\tau}_i \, [\varphi_i (\tau_k) -
    \varphi_i (\tau_{k+1})]}.
\end{equation}

Using the Poisson summation formula, the sum over angular momentum
configurations becomes a periodic sequence of narrow gaussians around
multiples of $2 \pi$
\begin{equation}
    \prod_i \, \sum_{m = - \infty}^{+
    \infty} \sqrt{\frac{2 \pi}{U \, \Delta \tau}} \, e^{- \frac{1}%
    {2 \, \Delta \tau \, U} [\varphi_i (\tau_k) -
    \varphi_i (\tau_{k+1}) - 2 \pi m]^2}
\end{equation}
that is the Villain approximation to
\begin{equation}
    \mathcal{T}_k \approx \prod_i \, e^{- \frac{1}%
    {U \, \Delta \tau} \cos [\varphi_i (\tau_k) -
    \varphi_i (\tau_{k+1})]}
\end{equation}
with dropped irrelevant prefactors.

What we get by means of this procedure is a mapping of the QPM into a
\emph{anisotropic} classical $(2+1)D XY$ model, with effective action
$\mathcal{S}$
\begin{eqnarray}
    \nonumber
    \mathcal{S} & = &
    K_{sp} \
    \sum_{
        \langle i,\, j \rangle , k
    }
    \left[ 1 - \cos \left(\varphi_{i,k} -
    \varphi_{j,k} -
    A_{i,\, j}\right) \right] \\
     & + &
    K_{\tau} \
    \sum_{
        i, \langle k,\, k' \rangle
    }
    \left[ 1 - \cos \left(\varphi_{i,k} -
    \varphi_{i,k'} \right) \right] \\
     & & K_{sp} = t \, \Delta \tau \hspace{1cm} K_{\tau} = \frac{1}%
    {U \, \Delta \tau}
\end{eqnarray}
where we used a symmetric notation for space and time lattice sites.
Since critical properties are not expected to depend on the asymmetry
of such model, and since for $\Delta\tau \To 0$ we have $ K_{sp} \To 0,
\ \	 K_{\tau} \To \infty \ \ \mbox{with} \ K_{sp} \, K_{\tau} =
\mbox{const.} $, one can fix $\Delta \tau = 1/\sqrt{t U}$. It then
follows that the coupling in the space and time directions are equal $
K_{sp} = K_{\tau} = K $. The isotropic model, Eq.(\ref{isotropic}) is
the one which is used in in the Monte Carlo simulations.

\bibliographystyle{prsty}

\begin{thebibliography}{99}
\bibitem{footnote3}
   In the perturbation expansion, the eigenfunction corresponding to
   the excited state is localized in space. Therefore one should not
   expect any condensation to a superfluid state.
\end{thebibliography}

\end{document}